\documentclass[rmp,aps,nofootinbib,sort,amsmath,amssymb,endfloats,preprint]{revtex4-1}

\def\gtwid{\mathrel{\raise.3ex\hbox{$>$\kern-.75em\lower1ex\hbox{$\sim$}}}}
\def\ltwid{\mathrel{\raise.3ex\hbox{$<$\kern-.75em\lower1ex\hbox{$\sim$}}}}

\def\spose#1{\hbox to 0pt{#1\hss}}
\def\simpropto{\mathrel{\spose{\lower 3pt\hbox{$\mathchar"218$}}
     \raise 2.0pt\hbox{$\propto$}}}

\newcommand{\araa}{{\it Annual Review of Astronomy and Astrophysics}}
\newcommand{\pasj}{{\it Publications of the Astronomical Society of Japan}}
\newcommand{\pasp}{{\it Publications of the Astronomical Society of the Pacific}}
\newcommand{\pasa}{{\it Publications Astronomical Society of Australia}}
\newcommand{\aj}{{\it The Astronomical Journal}}
\newcommand{\nar}{{\it New Astronomy Reviews}}
\newcommand{\mnras}{{\it Monthly Notices of the Royal Astronomical Society}}
\newcommand{\apjl}{{\it Astrophysical Journal Letters}}
\newcommand{\aap}{{\it Astronomy and Astrophysics}}
\newcommand{\apjs}{{\it Astrophysical Journal Supplement}}
\newcommand{\aaps}{{\it Astronomy and Astrophysics Supplement}}
\usepackage{graphics}
\usepackage{epsfig}
\usepackage{url}

\begin{document}


\input psfig.sty


\title{A panchromatic review of thermal and nonthermal active galactic nuclei}

\markboth{Antonucci}{Thermal and Nonthermal Active Galactic Nuclei}

\author{Robert Antonucci}
\affiliation{Department of Physics, University of California, Santa Barbara}


\begin{abstract}
The first short part of this review is a general, but very detailed, critique of
the literature advocating the existence of a class of Seyfert galaxies intrinsically
lacking broad emission lines and (sometimes) the thermal ``Big Blue Bump" optical
UV continuum component. Many uncertainties and erroneous assumptions are made in
these papers, and almost all the conclusions are weak or erroneous. Panchromatic
properties of all types of radio loud AGN are then reviewed in detail.

Radio galaxies usually show subparsec-scale radio core sources, jets, and a
pair of giant radio lobes. The optical 
spectra sometimes show only relatively weak lines of low-ionization ionic species,
and no clear nuclear continuum in the optical or UV region of the
spectrum. Some show strong high-ionization narrow lines. Finally, a few radio 
galaxies add broad bases onto the permitted lines. These spectral categories
are the same as those for radio-quiet AGN and quasars.

By the 1980s, data from optical polarization and statistics of the radio
properties required that many narrow line radio galaxies do in fact produce 
strong optical/UV continuum. This continuum and the broad line emission are
hidden from the line of sight by dusty, roughly toroidal gas distributions,
but they are seen in polarized flux.
{\it The radio galaxies with hidden quasars are referred to as ``thermal.''}
\par\vskip 2.5mm
\indent Do all radio galaxies harbor hidden quasars? We
now know the answer using arguments based on radio, infrared, optical and X-ray
properties. Near the top of the radio luminosity function, for FRII, GPS, and
CSS galaxies, the answer is yes. Below the top of the radio luminosity
function, many do not. At low radio luminosities, most do not. Instead these
{\it ``nonthermal'' weakly-accreting galaxies} manifest their energetic output
only as kinetic
energy in the form of synchrotron jets. These kinetic-energy-only (``nonthermal")
radio galaxies are a subset of those with only weak
low-ionization line emission. This applies to all types of radio galaxies, big FR II
doubles, as well as the small young GigaHertz-Peaked-Spectrum and Compact Steep
Spectrum sources. Only a few FR I sources are of the thermal type.
\par\vskip 10mm
\parbox{13cm}{
\begin{center}
{\bf To be published in\\ Astronomical and Astrophysical Transactions:\\ Journal of the Eurasian Astronomical Society}
\end{center}
\fbox{\fbox{\parbox{13cm}{\vskip 5mm
\begin{center}
\textbf{Ann. Rev. Astron. and Astrophy., rejected.}
\end{center}
\par\vskip 2.5mm
\begin{center}
\parbox[c]{12cm}{
Though the author was asked to write this review and spent much of a year doing it,
then revised it according to some of the input from the ARAA editors, the editors rejected
it. They demanded many ever-changing major and minor changes, many of which were described as
``non-negotiable,'' before they would reconsider it. The editors emphasized that they have no problems
with the scientific content, and that they had gone over the science ``with a fine-toothed comb.''
Nevertheless they may have found some errors if they had given the comb to an AGN person!
\par\vskip 2.5mm
The editors emphasized that they were rejecting the paper on purely non-scientific grounds.
I was informed for the first time after the rejection that the article had to be intelligible
to ``a beginning graduate student,'' in gross contradiction to statements in my invitation. That would
have destroyed the paper. A typical minor request: the editors urged me to take advantage of the large page margins to add
little comments and explanations which would better be put into footnotes. They apparently want it
to look like a chapter in an Astro 1 book.
\par\vskip 2.5mm
When ARAA calls you to write an article for them, remember that you may spend a year
on it and then just have it rejected unless you make major changes that you may not agree with.
If you write for them you could fall into this trap!}\vskip 5mm
\end{center}}}}}
\end{abstract}

\newpage

\maketitle
\tableofcontents
\newpage

\section{TERMINOLOGY}


Almost all Radio Quiet Seyfert galaxies, {\it defined by narrow emission line
ratios}, show the energetically dominant {\it thermal} ``Big Blue Bump" optical/UV
continuum components. Many are hidden from direct view, and are only seen in
polarized light. Many authors have argued, however, that some Seyfert 2s intrinsically
lack these components (or at least the broad emission lines) and such objects are
often referred to as True Seyfert 2s and sometimes as nonthermal Seyferts.  The
methodologies used are critiqued in detail, and it is concluded that all such claims
are dubious at best.

This paper concerns primarily radio loud active galactic nuclei (AGN). 
``Radio loud'' is sometimes defined by an absolute radio luminosity cutoff, and 
sometimes (less usefully) by a ratio of radio to optical luminosity. The nomenclature is
multifaceted, complex, and very confusing for a newcomer. We will divide 
the AGN into two broad classes, which correspond to the two popular and persuasive
central-engine models. The presence of an optical/UV continuum of the type
called the Big Blue Bump will be called ``thermal''
because there is a consensus that this is thermal radiation from a copious
opaque and probably usually geometrically thin accretion flow.  This 
includes the radio loud quasars, the Broad Line Radio Galaxies, and the objects that
have similar accretion flows hidden from the line of sight.
(Some papers define the Big Blue Bump as the excess over a
notional power law extending from the near-IR to the far-UV or X-ray, 
but that is not the most common usage, or the present usage.) The Big Blue Bump is
virtually always accompanied by conspicuous broad permitted emission
lines\footnote{\setlength\baselineskip{1ex}
Possible exception to this one discussed in the next 
section.} from regions collectively called the Broad Line Region. This combination is 
called a Type 1 spectrum.

By contrast, a radio loud AGN which lacks visible broad lines
is called a Narrow Line Radio Galaxy (``Type 2'' optical spectrum). The
narrow-line spectra of all radio types vary enormously from optically 
weak Low Ionization Galaxies --- sometimes loosely called Low Excitation
Galaxies --- like M87, to very powerful High Ionization Galaxies like 
Cygnus A. The former are turning out to be almost all ``nonthermal'' radio 
galaxies, lacking a powerful Big Blue Bump and Broad Line Region, even a hidden 
one.\footnote{\setlength\baselineskip{1ex}
Note that there is only a little evidence yet that the thermal and nonthermal
objects are bimodal in any property, and such isn't necessarily expected theoretically.}

\vfill\break

If the Big Blue Bump is directly visible in the total-flux spectrum, the object
is called a radio loud quasar,\footnote{\setlength\baselineskip{1ex}
Recall that this paper is largely
restricted to radio loud AGN.} or if low in optical luminosity
(e.g. $M(V) > -23$ for $H_0=50$ km sec$^{-1}$ Mpc$^{-1}$, as adopted 
for the older Veron-Cetty and Veron catalogs), it may be called a Broad Line Radio 
Galaxy.\footnote{\setlength\baselineskip{1ex}
It is sometimes argued that at these low luminosities, there are some relatively
subtle differences with respect to the quasars, e.g., \protect\citealt{2001A&A...379L..21V},
but such a distinction will not be made here.} In 
fact when it's clear from context, ``quasars'' will be taken to include Broad Line 
Radio Galaxies.\footnote{\setlength\baselineskip{1ex}
For  redshifts above a few tenths, the first two words of ``Narrow Line Radio Galaxy'' are 
often dropped, because the broad line objects are unambiguously called quasars.}

For radio-bright objects at redshifts of larger than a few tenths, the presence
of a (directly visible or hidden) optical/UV Big Blue Bump
is general --- except in those rare objects whose optical/UV spectrum is
overwhelmed by beamed synchrotron emission from the bases of favorably
oriented relativistic jets (``Blazars'').
For objects with a large contribution to the optical/UV continuum by
highly variable, highly polarized beamed synchrotron radiation, the
general term is Blazars, defined in \citealt{1978bllo.conf....1S}.
Blazars are defined as the union of two
classes: 1) objects in which a Big Blue Bump/Broad Line Region is still 
discernible against
a strong synchrotron component (Optically Violently Variable Quasars, also
known as Highly Polarized Quasars) with 1960s-1970s technology, and 2)
objects with a pure synchrotron continuum in those old spectra, and
little or no detectable line emission or absorption (BL Lac Objects).
However, it's been known since the 1970s at least that many
historically defined ``BL Lacs'' show emission lines, both narrow and
broad, especially (but not necessarily) when observed in low states. For
example, BL Lac itself has weak narrow emission lines, and stellar
absorption lines \citep{1981PASP...93..681M}; now we know
that broad lines are often visible as well \citep{1995ApJ...452L...5V}. In fact
it is well known that many highly polarized, violently variable quasars
are indistinguishable from BL Lacs when in high states
\citetext{\citealp{1978bllo.conf..228M}; see also \citealp{1981PASP...93..681M}}.

None of these, ``BL Lac'' nor ``High Polarization Quasar,'' nor
``Optically Violently Variable Quasar'' is very well defined, and many
studies have been damaged by blindly using these historical categories
--- sometimes just from catalog classifications, or by trying to mimic
them with equivalent width cutoffs, which result in classifications
changing with time \citep{1987AJ.....93..785A,2002apsp.conf..151A}! It is
also still often incorrectly asserted that the parent population
(equivalent objects at more than a few degrees inclination) for ``BL
Lacs'' is FR I double radio sources\footnote{\setlength\baselineskip{1ex}
FR I and II radio sources are
discussed in Sec.~\protect\ref{true-seyf2}.} (the lower-luminosity edge-darkened
ones). Complete disproof of this assertion has been around for decades,
e.g., \citet{1991BAAS...23R1420K}.
Yet people still write about FR II radio emission in BL Lacs as a
``problem'' for the unified model. 
This has invalidated studies of unification, cosmological evolution, and radio physics
\citetext{e.g., \citealp{1985Natur.318..446O,1990Natur.344...45O}}. See \citet{1999MNRAS.304..160J} 
for a well-informed and sensible discussion. Similarly many beamed radio quasars
have diffuse emissions similar to FRI galaxies, e.g. \citet{2002apsp.conf..151A}.

Great care is required in classifying AGN, and the price of carelessness is
spurious results. For example, historically 3CR382, 3CR390.3 and similar objects
were called Broad Line Radio Galaxies (Type 1 optical spectrum), and this is
still reasonable. But the same was done for 3CR234, because in fact the
broad H-$\alpha$ line is visible in the total-flux spectrum. This wasn't
an error in its historical context, but we now know that the broad
lines and Big Blue Bump are seen only by reflection \citep{1982Natur.299..605A, 
1984ApJ...278..499A,1995AJ....110.2597T,1998MNRAS.294..478Y}. Therefore from the point of view of
unified schemes, such objects must be included with the Narrow Line
Radio Galaxies (Type 2 optical spectrum), just as we keep Seyfert 2s (with
hidden Big Blue Bump/Broad Line Region) separate from the Seyfert 1s. It would
be great if a multi-dimensional classification scheme could be shown in a
drawing, but a confusing ``tesseract'' would be needed \citep{1993NYASA.688..311B}. 
Even a tesseract might oversimplify the true situation.

\section{Thermal and Nonthermal Emission in Seyfert Galaxies\label{sec:2}}
\subsection{AGN with Big Blue Bumps, but said to lack Broad Emission Lines}
Laor and Davis' (\citeyear{2011MNRAS.417..681L}) modeling paper has a convenient list of
observational papers reporting on a very, very few quasars with Big Blue Bumps,
yet little or no detectable broad line emission.  One limitation of these
analyses is that any equivalent width upper limits rely on making an assumption
about the broad line width. However, extremely large line widths are
exceptional for quasars, and do not occur in the faint but detectable broad
lines in some of these very weak-lined quasars, so the limits on broad line
equivalent widths are probably fairly close to the truth.

\citet{2011MNRAS.417..681L} also cite some
papers on a very different kind of AGN. These papers claim \citetext{dubiously in my opinion:
\citealp{2002ASPC..284..147A}} that for a few nearby AGN there is intrinsically
no detectable broad line emission, and in some cases no thermal Big Blue Bump,
at an interesting level. The authors refer to these as ``True Type 2
Seyfert Galaxies."

\subsection{Critique of polarimetric evidence said to support the existence
of True Seyfert 2s}
Most well-known Seyfert 2s have hidden broad line regions (HBLRs), often detectable
in polarized (scattered) light. Many authors assert that there are others which lack
BLRs, both visible and hidden. They may be called ``True Seyfert 2s," or synonymously
``Non-Hidden-BLR" (non-HBLR) Seyfert 2s. (Some authors seem to consider any lacking
the BBB/BLR component to be non-thermal emitters.) This has led to a small industry
which attempts to find other differences between the two types.  In my opinion,
most of the arguments for True Seyfert 2s are spurious and there are few if any strong
candidates, except the one mentioned at the end of Section \ref{sec:2}. Then the
secondary literature comparing the putative two types of Seyfert 2 has no merit.

\subsubsection{HBLRs revealed later}
Seyfert 2s are sometime said to lack broad lines in polarized flux, which is
undefined because it's a qualitative statement based on quantitative data
of limited SNR, and subject to interpretation. One should {\it never} say,
``certain objects lack a hidden BLR." It's crucial to say that they lack
a {\it detection} of an HBLR, or else the deductions made from it will be incorrect.

Hidden broad lines are often detected at later epochs, or at higher SNR,
so a classification of a non-HBLR nuclei today doesn't mean it will still
be one tomorrow. \citet{2010ApJ...711.1174T,2010ApJ...715.1591T} notes that:
``Changes in the central engine or the obscuration could result in
changes in appearance of the AGN. For example, NGC 2110 and NGC 5252
{\it were originally classified as non-HBLR S2s}\/, but perhaps due to their
intrinsic broad-line variability, later found to be [HBLRs]."  Another
reason broad lines may appear in later data in some cases is the vastly
improved SNR of (for example) Keck data relative to the original Lick
observations.

Finding reflected broad lines is {\it very} photon-intensive and you have to be
very lucky in finding a well-placed mirror with the right properties.

\subsubsection{Lack of stated upper limits}
Few if any upper limits have been published for broad lines of candidates,
in either flux or polarized flux. This is astonishing to me.

Flux limits would be more helpful because the limits in equivalent width
are very misleading unless they are quoted relative to the pure AGN continuum.
That's never done as far as I know, for the simple reason that the continua are
virtually pure starlight in most cases, and no AGN continuum can be discerned.
Upper limits to line equivalent widths must be raised by a very large and unknown
factor because of this effect, before comparison with Seyfert 1s.

\subsubsection{Polarization may not be due to scattering}
If there {\it are} actually upper limits in any published papers, they
wouldn't be intelligible
without showing that the polarization is caused by scattering of nuclear light.
This is done only intermittently for the True Seyfert 2 candidates.

Now let's consider the assumption of scattering polarization in more detail.  For a large
minority of Seyfert 2s, the nuclear polarization is due to dust transmission
on large scales, usually in the host galaxy plane.  Few authors seem to know this!
The arguments for my statement are 1) there is a dearth of edge-on Type 1 Seyferts
\citep{1980AJ.....85..198K,1982ApJ...256..410L}, and  2) the AGN axes are
unrelated to those of the hosts.  Thus many Seyfert 2s get their classification
from host-galaxy obscuration of the BBB and BLR sources.

But more to the point, they get their polarization from transmission through
large-scale dust, as shown definitively by \citet{1988ApJ...330..121T};
see also \citet{1999ApJ...525..673B} for the nature of polarization of low luminosity AGN.
While large-scale dust transmission polarization is usually in the disk plane
for Spirals, it can be in any direction in Ellipticals.

This effect has baffled most X-ray astronomers, who think that a Type 2
classification necessarily indicates a high-inclination view of a nuclear torus.

Objects with dust transmission polarization can be recognized by the shape
of their (Serkowski) polarization curves if the nuclear emission can be isolated
and the SNR is high.  Few (but some!) True 2 candidates have data
with sufficient SNR to test this possibility; that's needed because
this very common and mundane explanation should be ruled out before invoking
a more exotic scenario.

Also, it is by no means trivial to correct the data for Galactic interstellar
polarization.  A good simple recipe is given in the appendix of \citet{1984ApJ...278..499A};
it is adequate for most projects. A much more careful treatment is used in
\citet{2004MNRAS.354.1065K}.

As with all these points, I'm not implying that there are no good papers
which address these points as best as can be done. Perhaps my intended reader
is an astronomer who wants to read the papers critically, and concentrate
on the better ones.

To summarize, when the polarization is due to dust transmission rather than scattering,
which is very common, the polarized flux is roughly just a much noisier version
of the total flux and means nothing.

\subsubsection{Upper limits depend on assumed line widths}
If upper limits on equivalent widths or fluxes on broad lines
are (ever) presented, they would actually depend very sensitively on the assumed
line widths. This is what led \citet{1994AJ....108..414S} astray in
claiming (after a lot of Keck time doing spectroscopy) that the radio galaxy
Cygnus A lacks broad emission lines. Of course they had
to assume a particular line width in order to calculate their upper limits
to any possible broad emission lines.

The scattered lines just happen to be much broader than
\citeauthor{1994AJ....108..414S} assumed. If interested in
the gory details, see \citet{1990MNRAS.246..163T}; \citet{1994Natur.371..313A}; and
\citet{1997ApJ...482L..37O} for correction of the \citeauthor{1994AJ....108..414S} claim.

In propagating the errors in any argument, a large term needs to be included for this.

\subsubsection{``Secure non-detections" of broad lines}
\citet{2011ApJ...726L..21T} argue for a lack of BLR components in total or
polarized flux in NGC3147, NGC4698, and 1ES 1927+654. (I will return to NGC3147.)
The non-detections of polarized broad lines in all three is described as
``secure."  I can't find equivalent width or polarized flux limits for any of
them in the paper, however.

And outside of some very specific context, I'm not even sure that a
non-detection can {\it be} secure.

\subsubsection{Spectropolarimetry at H$\alpha$ often required}

At modest SNR, H$\alpha$ observations are required because it's a much stronger line 
than H$\beta$, and less subjection to extinction. Please see \citet[Fig.~2]{1984ApJ...278..499A}, 
and \citet{1994ASPC...54..201L}, where this is discussed in detail.
Usually limits on H$\beta$ are of only mild interest to me.

\subsubsection{Near-IR polarimetry often required}

Often the obscuration of the reflecting region is too great even to detect H$\alpha$
in polarized light.

In the nearest, prototypical radio galaxies Cen A (FR I) and Cyg A (FR II),
the host galaxies are very well-resolved. They famously have gorgeous
1--100kpc dust lanes with A(V)$\sim$ a few, which obscure the whole nuclear region.

In both these cases, the optical polarization is very low ($\ltwid1$\%),
but the NIR polarization is extremely high \citetext{$\sim20$\%: \citealp{2000ApJ...544..269C}
for Cen A; \citealp{2000MNRAS.313L..52T} for Cygnus A}.
The reason is that the nuclear occultation/reflection
regions are only visible through the dust lanes at relatively long wavelengths.
Barvainis and I predicted and found the same effect in 3C223.1, and in
that paper we discuss this situation generally \citep{1990ApJ...363L..17A}.
In all three cases the polarization angle is precisely perpendicular
to the radio jets, as expected in the Unified model. By the Copernican Principle,
it is likely that in other cases one would have to go even farther into the IR to
detect the occulation/reflection region, so these observations only provide
a lower limit to those lost to optical spectropolarimetry.

This is just one example of the incompleteness of optical polarization surveys
for hidden broad lines.

\subsubsection{Dilution of polarization by a starburst blue/UV component}
The nuclear regions of most well-studied Seyfert 2s are dominated by starburst light
in the blue/UV part of the spectrum. Many papers \citetext{\citealp{1995MNRAS.272..423C,
1995ApJ...452..549H}, etc.} have shown this.

In spectropolarimetry, the giveaway is higher polarization in
the broad lines than in the continuum ({\it after} careful subtraction of the old stars).
There is a ``second (blue) featureless continuum" source which dilutes the polarization in the
continuum but not in the broad lines; it is so nearly ubiquitous that \citet{1988ApJ...331..332G}
gave it a name, FC2; \citetext{see also \citealp{1995ApJ...440..597T}}. But FC2
is not featureless in the UV; it shows strong hot-star photospheric features.
\citetext{This was first discovered in \citealp{1984ApJ...278..499A}, reporting on 3C234, the
first reflected-light AGN.}

Similarly there is an HST survey of ultraviolet spectropolarimetry
of Seyfert 2s \citep{1996VA.....40..149H} which confirms the great strength of a
(low-polarization, starburst) continuum source dominating over the scattered nuclear
light in most cases.

Many papers use the observed continuum polarization directly,
although it has no meaning at all.

On a related note, it has always baffled me that \citet{1995ApJ...440..597T} has
reported some very low (2--3\% for Mrk 477) polarizations for broad emission lines in certain
Seyfert 2s. As explained in \citet{2002ASPC..284..147A}, ``the way to measure
the broad line polarization, and hence the polarization of the scattered light
alone, is to divide the polarized flux of the line by its total flux. However, it
is generally impossible to see the line clearly in total flux, so one just derives high
lower limits to the broad line polarization in most cases. \citet{1995ApJ...440..597T}
claimed to measure the broad line percent polarization, and thus needed the total
line flux in some objects. However, M.~Kishimoto and I looked carefully at the
case of Mrk477, in preparing a paper on polarization imaging. We examined the
total-flux plots, overlaying the permitted Balmer lines on top of various forbidden
lines; we saw little or no evidence for broad wings in the total flux. (You can
try this in your own homes.) We thus placed an upper limit on the total flux of
the broad components (with shaped matched to that of the lines in polarized flux),
leading to conservative lower limits on broad H$\alpha$ and H$\beta$
polarizations of 10\%. Thus we disagree with the intrinsic values of only 2--3\% quoted
in \citet{1995ApJ...440..597T}. H.~Tran kindly allowed us to use the same data
that he used.

\subsubsection{It takes luck to find hidden broad line regions!}
From \citet{2011ApJ...726L..21T}: ``Could the [True 2s] be the hidden counterparts
to the [NLS1]\dots? This can be ruled out by the simple observation that
no emission lines of any kind, broad or narrow, are seen in the polarized flux spectra."

That's absolutely wrong because you only get scattered broad lines when
you have a favorably placed mirror with significant optical depth and
covering factor. You have to be damn lucky and surely this is a case of
``absence of evidence does not equal evidence of absence."

\subsubsection{Isotropic selection is key for polarization and other surveys}
Several early papers reported that well-studied Seyfert 1s and 2s differ
systematically in various intrinisic properties. For example, it was claimed that
the Type 2s have stronger FIR emission, stronger CO, etc, and that this limited
the applicability of the spectroscopic aspect of the unified model. \citetext{The same
sorry story played out in tests of the beaming aspect of the unified model;
see e.g., \citealp{1985ApJ...294..158A,1983ApJ...271L...5H}.}

NGC1068 and NGC4151 have the same UV excess, but we're only seeing 1\%
of the nuclear UV in the former case, so that AGN comes from five magnitudes
higher on the luminosity function!! Nothing can be concluded about unification
by comparing NGC1068 to NGC4151.

More recently, \citet{2008MNRAS.385..195B} uncritically attribute to
\citet{2002ApJ...579..205G} the conclusion that Seyfert 2s with ``no hidden broad
lines" are more luminous than those with hidden broad lines. There are no
selection criteria at all, except that they need to have polarization
measurements in the literature! There is an obvious and huge selection effect
\citep{2002ASPC..284..147A} such that the hidden BLR objects tend to
have systematically larger optical flux, and even more importantly,
they have much better contrast against the host galaxy starlight. That's
probably why the hidden BLRs were detected in the first place.
\citetext{See \citealp{1994ApJ...430..196K} for near-proof!}

A very powerful comparison of the Seyfert types was made by 
\citet{1994A&A...283..791K}, who expertly analyzed the spectra of a substantial
sample of Seyferts of both types. The reason this is a fundamental paper which
is much, much stronger than all preceding and almost all subsequent analyses of
Seyfert galaxies is that the objects were selected independently of orientation with
respect to the line of sight. Thus the Type 1s and the Type 2s can be considered
equivalent in the unified model (modulo the statistical preference for 2s expected
for large torus covering factor, a modest effect). The samples were selected
by $60\mu$ flux, with a modest warmth criterion using $25\mu$ flux, which just
excluded most of the starburst galaxies.

We have followed up this sample with a rather long series of papers, e.g.
\citet{1999ApJ...526L...9P}, \citet{2001ApJ...555..663S}, and
\citet{2003ApJ...597..768S}, etc. As expected in the Unified Model, most or all
of the claimed intrinsic differences between Seyfert 1s and 2s disappeared.

Samples which aren't isotropically selected produce uninterpretable (actually
misinterpretable) results, because they compare 1s and 2s which are
intrinsically quite different.
Studies with highly biased samples have no merit.

Here is a related and shockingly widespread practice: Conclusions are routinely
drawn from noisy luminosity-luminosity plots: \citetext{e.g., the influential paper
\citealp{1991Natur.349..138R}}.
I'm sure that plotting the number of bars versus the number of bookstores in
cities and towns will show a fantastic correlation with a nearly infinite
statistical significance, but that doesn't mean drinkers like to
read, or that readers like to drink. Bigger cities just have more of everything.
Similarly plots involving dependent axes are ubiquitous in the journals, probably
because you always get a correlation if you do that. A higher level of scholarship
will be required to make progress.

As noted, there is now a second-order body of spurious literature resulting
from the True 2s movement: there are many other papers which supposedly compare
the hidden broad line objects with those without detections in existing data,
without due (or any) regard for these concern listed above. In my opinion, this
literature has no merit.

\subsection{Critique of non-polarimetric evidence said to support the existence of True Seyfert 2s\label{true-seyf2}}
\subsubsection{Independent analysis by \protect\citet{2012arXiv1207.5543S}}
\citet{2012arXiv1207.5543S} have stepped in to provide the upper limits on broad
line emission for the putative True 2s (including NGC3147) which are conspicuously
absent in most or all papers proposing candidates. They assumed an uncertain but
reasonable velocity prescription for determining the limits to broad emission
lines in ``True 2s," and presented the relationship of their upper limits,
normalized to hard X-rays. What fraction of Seyfert 1s in
their sample had a value for this that's less than their upper limit for
NGC3147? Is it very small? No, it's 50\%! They did, however, find that two
other objects have ratios less than that of almost all of
their Seyfert 1s (their Fig.~14). Remember though that their upper limits refer
only to lines matching their velocity prescription, so they are really lower
limits not upper limits!

\subsubsection{Rapid X-ray variability}
A key to distinguishing an ordinary very common Narrow Line X-ray Galaxy from a
scattered source is {\it rapid} X-ray variability. \citet{2008MNRAS.385..195B}
show a variation in X-ray flux in their favorite candidate NGC3147 over an
interval of 3.5 years, arguing that the change favors the interpretation that
the X-ray continuum is seen directly like that in Seyfert 1 nuclei. But as noted
above, much more rapid variations (weeks to months) have been seen in the scattered
broad lines in several objects, so the scattering region can easily be small
enough to accommodate the X-ray variation in these candidates, which have
(almost?) invariably very low luminosities.

In the improved follow-up by \citet{2012A&A...540A.111M},
their Fig.~1 provides good information on short term x-ray variability
over $\sim$200,000s.

There isn't any.

Please compare with the behavior of some other (non-cherry-picked)
exposed nuclei (Seyfert 1s). Figure~1 of \citet{2005MNRAS.363..586U} show a
similar monitoring period for NGC3227 and NGC5506; and Fig.~1 of \citet{2008A&A...490..995P}
has four others. The lack of short-term variability in NGC3147 seems to be atypical and perhaps
unique if it's an exposed Seyfert 1 type X-ray source. The very low luminosity
of the True 2 candidates suggests particularly rapid variability
should be seen if the X-ray sources are directly exposed.

Conclusion: though the X-ray change in 3.5 years in NGC3147 was touted
as favorable to the idea of an exposed Seyfert 1-like nucleus, the lack of
short-term variations show that just the opposite is the case.

Note: there is a new paper with simultaneous X-ray and optical data, 
by \citet{2012arXiv1209.0274B}, which I haven't yet studied carefully. According to the abstract, 
out of the eight candidates from the literature which they have studied, 
they have debunked four. For the remaining four, they favor a True 2 interpretation 
based on non-detections of broad H$\beta$. Many of the points above should be considered 
when judging this evidence for True 2s. I believe they negate the argument.

\subsubsection{Do low X-ray columns indicate a clear optical line of sight?}
Low x-ray columns are often invoked legitimately in arguing for an exposed
nucleus, if the source can be shown to be Compton thin (strong soft X-rays),
and if A(V) / N(H) has the value for diffuse Galactic interstellar
clouds. In fact, AGN torus dust tends to have less optical absorption and
reddening than standard dust, e.g. \citet{2001A&A...365...28M}, strengthening this argument.

Of course one can measure a spurious zero column for opaque objects like NGC1068
because the X-rays, like the optical, are only seen in scattered light. However,
X-ray astronomers are usually careful to consider this possibility, and often
argue persuasively against it.

In a few cases, e.g. \citet{2010ApJ...722...96S}, it has been found that the
soft X-rays arise from off the nucleus, from diffuse emission or interlopers,
invalidating the low column measurement. Presumably these situations are rare.

\subsubsection{Near-IR emission lines}
If the absorption is relatively low, as in a Narrow Line X-ray Galaxy (a well-studied class
popular in the 1980s), in principle one could see the broad lines in the NIR
\citetext{or even weakly at H$\alpha$: \citealp{1980ApJ...240...32S}}. It seems that
published arguments are based at least notionally on observed upper limits to
equivalent widths, though I'm not aware of any quantitative values.
These would need to be corrected carefully because the NIR is strongly
starlight-dominated in Seyfert 2s, especially in the True 2 candidates. There
is often a hot dust contribution as well. The corrected upper limits, that is,
the values relative to the AGN continuum, would come out very much higher than
those measured, and extremely uncertain.

Advocates could make stronger arguments with fluxes rather than equivalent widths,
e.g. by setting limits on the flux of a NIR broad line relative to x-rays. One
still has to worry about the line width, but avoids the perils of {\it drastic}
equivalent width dilution as described above. The host galaxy dominates
in virtually all Seyfert 2s, and it dominates very greatly in the True
2 candidates.

\subsubsection{The mid-IR calorimeter}
An excellent test for a hidden AGN is the ``mid-IR (warm) calorimeter," which
indicates the luminosity of almost any visible or hidden AGN approximately but
robustly. That's why many of us have worked very hard on high-resolution
mid-IR imaging \citetext{\citealp{2004ApJ...602..116W} and others cited later in
this review}.

More importantly, there are many papers by Ogle's group and others on large
surveys with the Spitzer IRS spectrograph. Ours and some of the others are
based on nearly isotropic samples, i.e. they select on radio lobe flux.
Hopefully, my citations later in this review are nearly complete to early 2011.

Clearly, this parameter must be considered when advocating hidden AGN,
or the lack thereof.

\subsubsection{Are the True 2s even Seyfert 2s at all?}
Congratulations to any readers who have gotten far enough to reach this key point.

Some of the True 2 candidates have optical spectra inadequate to support
a Seyfert classification, or even adequate to disprove a Seyfert classification.
I think a True Seyfert 2 should have to be a Seyfert 2, but several proponents
disagree, at least implicitly.

As noted, according to \citet{2012A&A...540A.111M}, the very best candidate
may be NGC3147, so I looked at the optical spectra. According to
\citet{2010ApJ...722...96S} and pc's from the authors, the best data for line
measurements is the spectrum in \citet{1995ApJS...98..477H}. The main reasons
is that the line measurements in this case were made after host starlight
subtraction. Please take a look at their Fig.~27.
The line measurements are given in \citet{1997ApJS..112..391H}. For [O III]
$\lambda$5007/H$\beta$, a value of 6.1 is given, with moderate error bars as
discussed in the paper. The narrow H$\beta$ line cannot be seen at all in the
unsubtracted spectrum! Thus one would na\"\i{}vely set a high lower limit
to the [O III]$\lambda$5007 / narrow H$\beta$ ratio. But this is entirely
incorrect because as \citet{1995ApJS...98..477H} state (though the figure is
not shown), after careful host subtraction, the narrow H$\beta$ line is clearly
detected! Now look at [N II]$\lambda$6584/H$\alpha$: the measurement is given
as 2.71, which is past the Seyfert region in some diagnostic diagrams such as
those of \citet{1987ApJS...63..295V} and \citet[Fig.~1]{2012arXiv1209.0274B},
though those boundaries are somewhat subjective.

The candidate IRAS01072+4954 has an H-II region spectrum! See \citet{2012A&A...544A.129V}.

The [O III]$\lambda$ / narrow H$\beta$  and [N II]$\lambda$6584/H$\alpha$ ratios 
are biased upward in the uncorrected spectrum. The reason is that 
the Balmer lines, but not the forbidden lines, are sitting in 
stellar absorption lines. Thus the uncorrected measurements have no meaning, 
and the results depend mainly on the accuracy of the starlight subtraction 
(assuming the authors always do it!). Of course this is a very serious 
problem for low-luminosity AGN, like most of the True 2 candidates,
and it's not a serious problem at higher AGN luminosities. 

A good quantitative feeling for the size of the effect can be found in
\citet[Equations 2, 3 and 4]{1983ApJ...269..466K}. Naturally it depends on the
age of the stellar population, parametrized by \citeauthor{1983ApJ...269..466K}
in terms of continuum colors.

I think that many LINERs lack the BBB/BLR at an interesting level, as described
in great detail in this paper in the context of radio-loud AGN. However as
documented ahead, a significant minority {\it do} show thermal MIR reprocessing
bumps however, and more importantly many {\it do} show broad lines in polarized
flux \citetext{e.g. \citealp{1999ApJ...525..673B}}.

Seyferts were always and remain defined solely by optical spectral
line ratios, and to show that an object is a True 2 means first to show
that it is a Seyfert galaxy to begin with.

A True 2 also means (to me) that despite a secure Type 2 classification, there are upper
limits (with all the many caveats described above) to the broad line equivalent
widths which are well below some adopted normal range for objects selected in
a way to make them as analogous as possible to the candidate. The equivalent
width limits {\it can't} be given relative to the host starlight, or in the NIR,
relative to the host and dust emission. It must be relative to the Big Blue
Bump component, or it means nothing at all.

\subsection{Disclaimers}
\begin{itemize}
  \item Some of what I said is probably wrong.

  \item I'm critiquing. I haven't given all the arguments in favor of True 2s,
  but I haven't cherry-picked: I mentioned the arguments that seem important to
	me at this moment. Unlike Fox News, I do not strive to be fair and balanced.

  \item Not all of my arguments apply to all objects or all papers!! These are
	mainly general comments that I consider in forming an opinion on True 2 candidates,
	and I heartily recommend that others consider them also.
	I do {\it not} want to tarnish the whole field, but I'm afraid I may have
	made some authors feel guilt by association. The reader must not generalize.

  \item Some of the points argue the plausibility of identification as common
	NLXGs, and some of hidden BBB/BLRs. I'm not trying to present a single scenario
	for all --- objects must be considered very carefully and individually. And for
	some I may mention arguments for both scenarios. I'm not striving here for a
	big picture.

  \item I don't really give a damn about whether or not there are True 2s.

  \item Finally, I know of one excellent candidate for a True 2 in a radio
  galaxy: see \citet{1997ApJ...484..193C}!!  This high-ionization object is a
	bit complicated but the arguments given are persuasive.

\end{itemize}

\section{BROADEST SKETCH}
\subsection{Where To Find Basic Information}
Active Galactic Nuclei (AGN) is a term encompassing a variety of energetic
phenomena in galactic centers which are thought to be powered directly or
indirectly by accretion of matter onto central supermassive black holes. 
There are many reviews of this field, including several of book length.
Perhaps the most technical and broad book-length review is that of
\citeauthor{1999agnc.book.....K}, which was published in \citeyear{1999agnc.book.....K}.
The classic text on spectral analysis of gaseous nebulae
and AGN has been updated in 2006 \citep{2006agna.book.....O}; it's a wonderful
book, but it is biased towards the optical and ultraviolet regions of the
spectrum. A new edition (2009) of An Introduction to Radio Astronomy by 
\citeauthor{2009ira..book.....B} is also noteworthy.

Few theoretical predictions have been borne out in this field, and 
understanding is still semi-quantitative at best \citep{1985ApJ...288..205A,1988smbh.proc...26A, 
2002apsp.conf..151A,1991A&A...248..389C,1999PASP..111....1K,2007ASPC..373...75B}. This 
review will refer to some fairly general theoretical ideas, but it will mainly 
organize some observational information on radio galaxy central engines that has 
become clear over recent years. It will not include a general introduction to 
AGN (see above references), but will address the nature of the central engines in 
radio galaxies which is a topic tied up as a practical matter with orientation 
effects on observations. Orientation effects are introduced in the next 
section, and as needed throughout the text.

The general topic of orientation effects (``Unified Models'') is reviewed in
detail in \citet{1993ARA&A..31..473A}. The material in that review is almost entirely
``still true.''  However, it has been updated and elucidated in
several more recent (but generally narrower) reviews \citep{1995PASP..107..803U,
1997PASA...14..230D,1999AJ....118.1963C,1999ASPC..162..101W,2001ASPC..249..193A,2008NewAR..52..227T}.

In a nutshell, prior to the mid-1980s, it seemed that radio loud quasars 
(with their powerful thermal optical/UV light) were quite distinct from radio
galaxies; the latter are analogous to quasars in general radio properties, but
apparently lacked the strong optical/UV electromagnetic luminosity (Big Blue Bump).
In a landmark review of bright extragalactic radio sources, \citet{1984RvMP...56..255B}
posited that {\it ``The ratio of mass accretion rate to the mass of the hole
may determine whether a [radio loud] active galactic nucleus will be primarily
a thermal emitter like an optical quasar or a nonthermal object like a radio galaxy.''}

Through optical spectropolarimetry and other means, it was subsequently
determined that many radio galaxies, especially the most powerful ones,
with strong high-ionization narrow emission lines, actually harbor hidden
quasars surrounded by opaque dusty tori, so that the
observational appearance depends on the inclination of the radio axis to the
line of sight. But now we know that for many radio galaxies, any hidden 
quasar must be very weak.

The radio quiet and radio loud objects with high ionization are 
virtually all visible (Type 1) or hidden (Type 2) Seyferts or quasars
\citetext{e.g., \citealp{2002ASPC..284..147A}}. At lower radio luminosities
of all radio types, we find mostly LINERs\footnote{\setlength\baselineskip{1ex}
A few LINERs have strong broad lines, e.g., \protect\citet{1984ApJ...285..458F}. 
Many LINERs have very inconspicuous broad H$\alpha$ components, but it isn't clear 
to me that they are strictly analogous to those in low-luminosity Seyferts. For 
example, we do not know whether they vary rapidly (L.~Ho, 2011, private communication).}
(Low Ionization Nuclear Emission Regions). A recent and comprehensive review of
LINERs is that of \citet{2008ARA&A..46..475H}.

\subsection{Nature of Geometrical Unified Models}
In the low-redshift universe, there is a near-perfect correspondence between
``radio loud'' objects --- with $L_\nu$(1.5 GHz) loosely extending from perhaps
$\sim10^{28}$--$10^{36}$ erg sec$^{-1}$ Hz$^{-1}$ --- and elliptical hosts. (This
paper uses H$_0=70$ km sec$^{-1}$ Mpc$^{-1}$, $\Omega_{\rm matter} = 0.3$, and
$\Omega_\Lambda = 0.7$.) For a fiducial $L_\nu\simpropto\sim\nu^{-1}$ spectrum,
for which there is equal power per logarithmic frequency interval --- also
referred to as ``per dex'' --- the parameter $\nu L_\nu$ gives the power
integrated over an interval of 0.30 dex (powers of 10). Thus if one integrates
such a spectrum over the ``radio region'' 30 MHz--300 GHz, the corresponding
luminosity is 13.3 times as great; a look at various radio AGN in the NASA
Extragalactic Database (http://nedwww.ipac.caltech.edu) shows that the
luminosity per dex of the dominant optically thin synchrotron component tends
to be lower outside this range, so depending on the application, one may
consider this as a crude ``radio bolometric correction.'' Resulting radio powers
can exceed $1\times10^{45}$erg/s. Estimates of energy tied up in 100-kpc scale
radio lobes are as high as $1\times10^{61}$ ergs or more, even using the
particle/magnetic field minimum-energy assumption and assuming a lack of a
dominant proton contribution. (The minimum-energy assumption posits that energy
is apportioned between relativistic electrons and magnetic field in such a way
as to minimize lobe energy content for a given synchrotron luminosity.)

Unified models assert that certain AGN classes differ only in orientation with
respect to the line of sight. These models comprise two separate (though
interacting) assertions, as illustrated in Fig.~1. The first to be recognized
historically is the effect of relativistic beaming (aberration causing
anisotropy in the observed frame) in the powerful synchrotron jets which feed
particles and magnetic energy into the radio lobes. When seen at low
inclinations, beaming amplifies and speeds up ``core'' (subparsec scale jet,
usual synchrotron-self-absorbed)
radio flux variability and (apparent faster-than-light) ``superluminal
motion.'' Thus a special fortuitous orientation of a nearly axisymmetric object
leads to a different observational category (Blazars). The same objects, seen
at higher inclination, are ordinary radio-loud galaxies and quasars
\citetext{\citealp{1984RvMP...56..255B,1985ApJ...294..158A,1992AJ....104.1687K},
etc.}. We can call this the beaming unified model (or more loosely, just the
beam model, though that term usually connotes some connection with actual physics).

At low redshift, radio-quiet (but not silent) AGN lie in spiral hosts, and go
by the name of Seyfert galaxies. They rarely show detectable motions in their
weak radio jets --- and when they do show motions, the apparent speed is usually
much less than the speed of light. More luminous radio quiet objects are called
Radio Quiet Quasars, or historically, Quasistellar Objects (QSOs).

\begin{minipage}{5in}
\vskip 4.5truein
\includegraphics{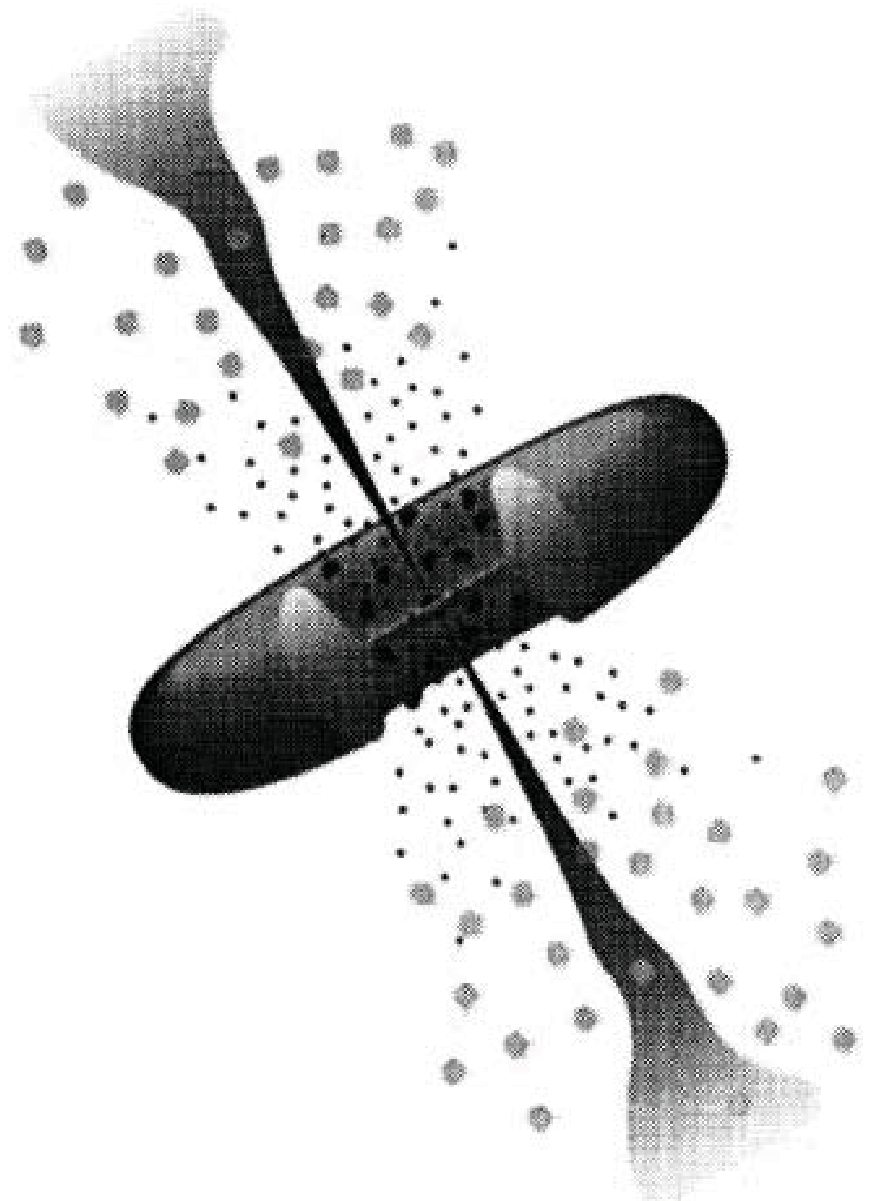}
\setlength{\baselineskip}{1ex}
\vskip 4mm
{Figure 1. A schematic diagram of the current paradigm for radio-loud AGN (not
to scale). Surrounding the central black hole is a luminous accretion disk.
Broad emission lines are produced in clouds orbiting outside the disk and
perhaps by the disk itself. A thick dusty torus (or warped disk) obscures the
Broad-Line Region from transverse lines of sight; some continuum and broad-line
emission can be scattered into those lines of sight by warm electrons or dust
that are outside the observing torus. Narrow emission lines are produced in
clouds much farther from the central source. Radio jets, shown here as the
diffuse 2-sided jets characteristic of low-luminosity, or FR I-type, radio
sources, emanate from the region near the black hole, initially at relativistic
speeds. \citetext{Adapted from \protect\citealp{1995PASP..107..803U}}}
\end{minipage}
\vskip 2.5mm

But both radio quiet and many radio loud AGN widely exhibit another kind of
orientation unification: many well-studied objects include energetically
dominant continuum components in the optical-ultraviolet region, referred to as
the Big Blue Bump, and widely attributed to thermal radiation from optical thick
accretion flows. \citetext{Confirmation of the latter can be found in
\citealp{2004MNRAS.354.1065K,2005MNRAS.364..640K,2008Natur.454..492K}.}
They are almost always accompanied by broad
(5,000--10,000 km/s) permitted emission lines. Both these components reside inside
optically opaque dusty structures which to zeroth order have the shadowing
properties of tori, and these structures are referred to loosely as the ``AGN
torus.'' In some cases we have a direct (low-inclination, polar) view of these
compact components (in quasars, Seyfert 1 galaxies, and Broad Line
Radio Galaxies). In many other cases (high inclination, equatorial views), we
can be sure these two components are present inside the tori using the
technique of optical spectropolarimetry, which literally allows us to see the
nuclei ``from above,'' using ambient gas and dust as natural periscopic mirrors.

This wonderful trick of optical spectrapolarimetry \citetext{e.g. \citealp{1982Natur.299..605A,1983Natur.303..158A,
1984ApJ...278..499A,1985ApJ...297..621A}} uses the polarization property of scattered
light to separate the spectrum of hidden sources from any sources of direct
light. The above references show that the Big Blue Bump and the Broad Line
Region are present but hidden from direct view. The scattering polarization
position angle indicates that the photons emerging from these components can
only escape from the hidden nuclei if moving along the radio jet (and lobe)
axis, so that the other directions must be blocked by an obscuring 
torus. This is illustrated in Fig.~2, from \citet{1995ApJ...440..597T},
illustrating the total and polarized fluxes, for 3CR234. The polarized photons
(scattered light) contribute to both plots, of course, but show up with great
contrast when plotted alone.

\begin{minipage}{5in}
\vskip 2.5truein
\includegraphics{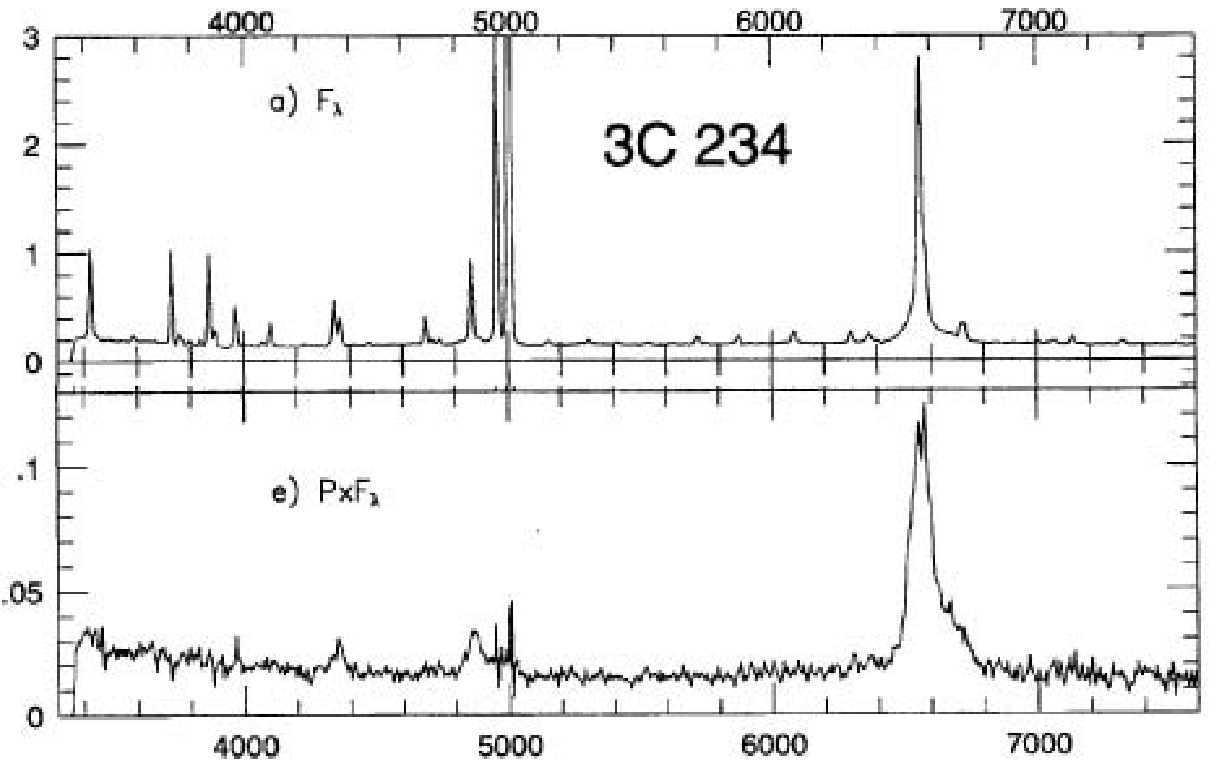}
\setlength{\baselineskip}{1ex}
\vskip 4mm
{Figure~2. Total and polarized flux for the Narrow Line Radio Galaxy that showed
as the first hidden quasar, 3CR234 \citetext{\protect\citealp{1984ApJ...278..499A}; the
figure shows the data of \protect\citealp{1995AJ....110.2597T}}. The polarization
angle relative to the radio axis shows that photons can only stream out of the
nucleus in the polar directions. The polarized flux (akin to scattered light)
spectrum shows the hidden quasar features at good contrast. (In this case,
however, the scattered ray is itself reddened.)}
\end{minipage}
\vskip 2.5mm

If all objects fit in with the above descriptions of orientation unification, we
would be left with as few as two independent types of ``central engines'' on
relativistic scales\footnote{\setlength\baselineskip{1ex}
This refers to regions where the jet velocity is
almost the speed of light, $c$, and also the gravitational potential energy per
unit mass is of order $c^2$.}, those for radio
quiet and those for radio loud AGN.  But we now know that many radio galaxies
lack powerful visible or hidden Type 1 engines (Big Blue Bump, Broad Line
Region, copious accretion flows). They occur frequently in some regions of
parameter space (as delineated for example by redshift and radio flux or
luminosity), and not in others. The purpose of this review is to gather the
multiwavelength evidence for the last two statements, and emphasize the
dependence of the distribution of central engine types on parameter space,
which is the key to avoiding many errors and much confusion and even discord.
We will see that a great deal of self-consistent information is available on
the occurrence of the two types of radio galaxy/quasar central engines.

\subsection{Types of Powerful Radio Source; Scope for Unification by Orientation}
Within the radio loud AGN arena, several types are also distinguished based not
on central engine properties but on extended radio morphology. Unfortunately,
they do not correspond very closely to the optical catagories!

The most luminous giant ($\ell\gtwid100$ kpc) double radio sources are
designated FR II (``Classical Double''), for \citet{1974MNRAS.167P..31F}; those
authors noted that a whole suite of properties change together fairly suddenly
over a critical radio luminosity (referring to the roughly isotropic diffuse
emission). Sources above $\sim2\times10^{32}$ erg/sec/Hz at 1.5GHz show
edge-brightening and hot spots, where the radio jets impinge on an external
medium, and shocks partially convert bulk kinetic energy to particle and field
energy. They also tend to have strong side-to-side asymmetry (generally
attributed to relativistic beaming) of the jets over scales from the relativistic
region up to tens of kpc. The lower luminosity giant (also $\ell\gtwid100$ kpc)
objects (FR I galaxies) also have strong side-to-side jet asymmetry, but only on
1--1000pc scales in most cases. An important refinement to the FR classification
scheme is that the dependence of the exact radio luminosity cutoff depends on
the optical luminosity of the host galaxy \citetext{\citealp{1994ASPC...54..319O}, but see 
\citealp[Fig.~4a]{2009AN....330..184B}}.

The ``FR'' types of radio galaxy generally have sizes of 25--1000kpc, but large
populations of smaller sources exist, and they can still be very powerful. (The
small sizes mean that their lobe energy content is very much lower, however.)
They are denoted in an inconsistent way, according to their means of discovery.
Optically thin, steep-spectrum radio sources which were historically unresolved
on arcminute scales, were (and are) called Compact Steep Spectrum
(CSS) sources to distinguish them from opaque beamed synchrotron cores; their
spectra peak at $\sim100$MHz, and they are generally defined to be in the size range
1kpc--15 or 25kpc. Sometimes they are crudely defined to be sources smaller
than typical host galaxies. Sources whose extent is less than 
$\sim1$kpc, with even more compact substructure, are
often dominated by synchrotron components which are self-absorbed up to $\sim$GHz
frequencies\footnote{\setlength\baselineskip{1ex}
It is proposed by \protect\citet{1999mdrg.conf..173B} that
the weak fluxes at low frequencies result from free-free rather than inverse
synchrotron absorption. Encouraging follow-up work can be found in
\protect\citet{2008ApJ...680..911S} and \protect\citet{2010ApJ...715.1071O}.},
and are called Gigahertz-Peaked-Spectrum sources (GPS). (Even
tinier sources are being sought by selecting for self-absorption peaks at even
higher frequencies.) These classes are reviewed by \citet{1998PASP..110..493O}. Since
that paper was written, much evidence has accumulated from VLBI proper motions
that the sources are small because they are very young ($\sim1000$--100,000
years!\footnote{\setlength\baselineskip{1ex}
This material was reviewed recently by \protect\citet{2008ASPC..386..176G}.}).
In at least some cases it is known from faint extended emission that these very
young ages refer only to a recent phase of activity, however. Statistically,
only a very small fraction of small and very short-lived sources can grow to be
huge bright long-lived sources.

The relevant properties of classes will be discussed in turn, generally starting
with the radio data and proceeding upwards in frequency. We will discover that
some FR II radio galaxies at $z\ltwid0.5$--1.0 lack powerful hidden quasars.
These objects {\it may} have hidden Type 1 nuclei, but they are constrained
to be much weaker than those of the ``matched''\footnote{\setlength\baselineskip{1ex}
In radio flux and redshift.} visible quasars, and thus they
are not ``unified'' (identified) with them via orientation with respect to the
line of sight. At low redshifts ($z<\sim0.5$) it is probable that only a minority
of FR II radio galaxies in the 3CR catalog host hidden quasars.

Next, we will tackle the less powerful (FR I) giant radio galaxies, which are by
selection nearby in almost all cases. The 3CR catalog flux cutoff of 10 Jy at
178MHz \citep{1983MNRAS.204..151L} corresponds to the FR I vs.\ II radio
luminosity separation at $z \sim0.2$. Most (but by no means all) of these
objects have nuclear spectral energy distributions dominated by synchrotron
radiation, with no evidence for visible or hidden ``Type 1'' central engines.

Finally the small, young GPS and CSS sources will be discussed. Very recent
information from the ISO and Spitzer infrared satellites has greatly increased
our knowledge of ``shadowing unification''\footnote{\setlength\baselineskip{1ex}
As a reminder, unification
of broad line and narrow line AGN by orientation of a toroidal nuclear obscurer
is here called ``shadowing unification.'' Ascription of superluminal motion and
other relativisitic effects in subpopulations to orientation is called ``beaming
unification.''} at various radio luminosities.

A major caveat of this paper, and of this field, is that most of the
information derives from the brightest radio sources, especially those in the
3CR catalog, so no implication is made for unexplored regions of parameter
space! Another major caveat is that while we discuss thermal vs.\ nonthermal
galaxies\footnote{\setlength\baselineskip{1ex}
Recall that thermal vs.\ non-thermal refers not to the radio
emission itself, but to the presence of an energetically dominant optical/UV
source thought to arise from an accretion disk.}, relatively little evidence is
presented that any parameter is bimodal, so that there could be a continuum of
properties.

\subsection{The Infrared Calorimeter\label{sec:broadinfraredcal}}
Radiation absorbed by the dusty torus is largely reradiated as infrared, and
many studies have concluded that in reasonably luminous AGN (so that the IR is
not dominated by a normal host galaxy), at least the near- and mid-infrared
reradiation ($\approx1$--40 microns) is dominated by reprocessed nuclear
optical/UV/X-ray light. In specific populations, the entire IR seems to be
radiated by the torus, because the colors are warm throughout, and there is
evidence for only weak star formation (e.g. PAHs) or synchrotron radiation
(e.g., strong radio-mm emission). Thus the infrared reradiation of nuclear light
can potentially be used as a {\it calorimetric indicator} for the
luminosity of any hidden AGN.

There are at least two ambiguities in using this infrared emission as an AGN
calorimeter. The first is that some of the nuclear radiation may not reach the
torus, either because of the intrinsic latitude-dependence expected for many
models of the Big Blue Bump \citetext{e.g. \citealp{1985MNRAS.216...63N}}, or preferentially
planar Broad Line Region absorption \citep{2001A&A...375...25M,2007arXiv0711.1025G}.
Both of these effects add noise to the dust reradiation calorimeter, and tend to
make hidden quasars look dimmer than visible quasars in the infrared for a given
opening angle. Nevertheless most studies show infrared luminosity about as expected,
from detailed studies of individual objects \citep{1984ApJ...284..523C,
1992ApJ...395L..73S} and for populations of Type 1 and Type 2 objects within
isotropically selected samples \citep{1994A&A...283..791K}.
Carleton et al.\ \citetext{\citeyear[see Fig.~3]{1984ApJ...284..523C}} shows how the
infrared calorimeter works, based on the first spectropolarimetric hidden AGN, 3CR234.

\vskip 5mm
\begin{minipage}{5in}
\vskip 2.5truein
\includegraphics{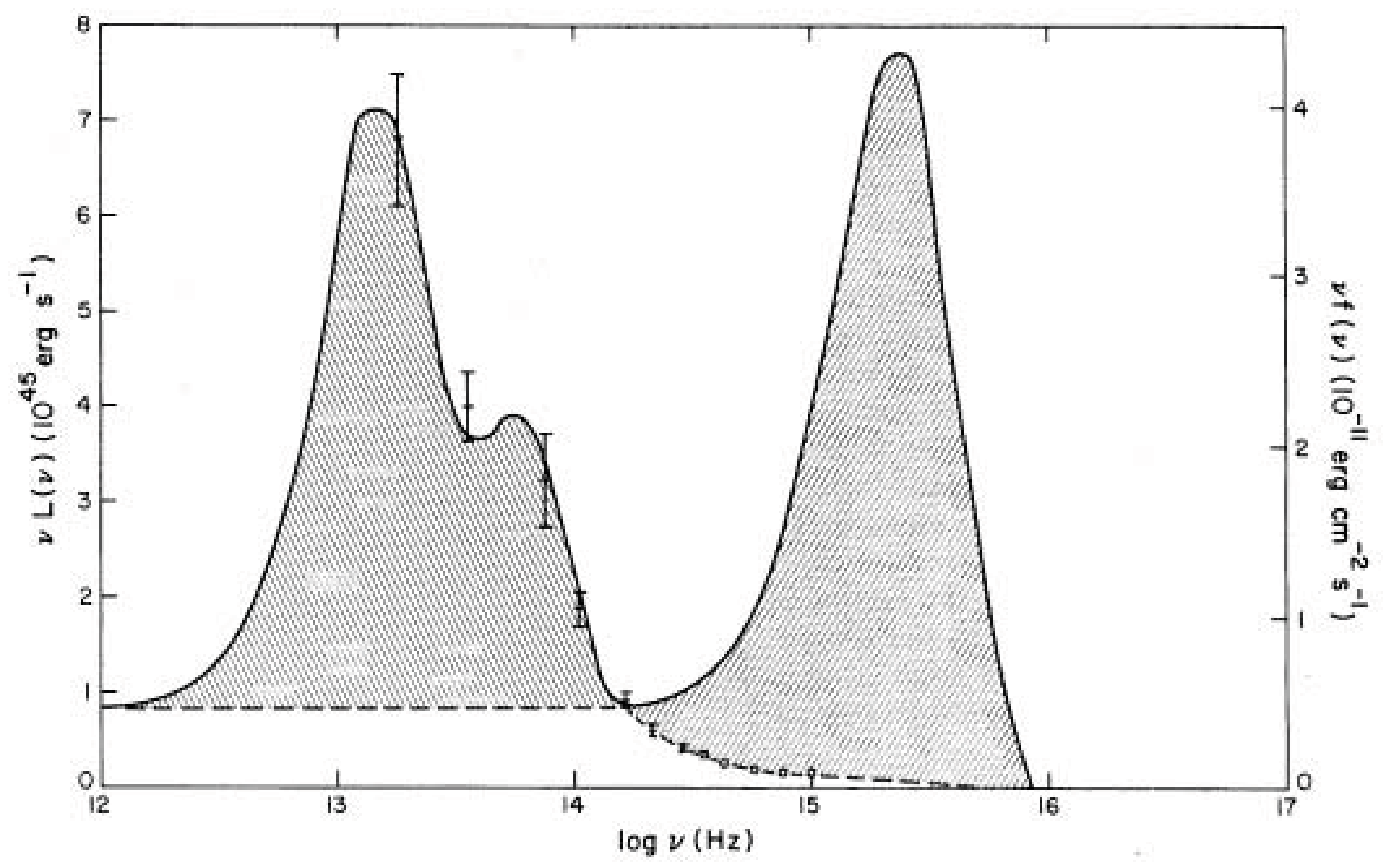}
\setlength{\baselineskip}{1ex}
\vskip 4mm
{Figure~3. 3CR234 continuum observations plus model fits. The ordinate is linear
in $\nu L_\nu$, and the area under any portion of a curve is proportional to the
luminosity in that portion of the spectrum. The hidden optical/UV component
(Big Blue Bump) has been reprocessed into the infrared in this hidden quasar.}
\end{minipage}
\vskip 5mm

Also, although the covering factor deduced from dividing infrared re-emission
by Big Blue Bump luminosities are therefore lower limits, they tend to be high
($\sim0.1$--1) so that they can't be too far off!
Although \citet{2007A&A...468..979M} didn't actually integrate over the SED and
should therefore be viewed with much caution, it is interesting that they
actually find some covering factors nominally greater than one for low-moderate
luminosity objects. This was predicted qualitatively by \citet{2004ApJ...616..147G},
who argued that a preponderance of large grains leads to less apparent absorption (spectral
curvature) in the optical/UV region than otherwise, and thus one may fall into
the trap of inferring very low extinction. The inner torus is expected to lack
small grains, most robustly perhaps on the grounds that the torus sublimation
radius must be larger for small grains according to radiative equilibrium. The
lack of small grains at the sublimation radius is supported by the data of
\citet{2006ApJ...639...46S}: see discussion in \cite{2007A&A...476..713K}.

Another concern with the infrared calorimeter is the expected anisotropy of the
thermal dust emission due to the large dust column densities
\citep{1992ApJ...401...99P,1993ApJ...418..673P}.
Many Type 2 AGN have X-ray columns of $\gtwid 1 \times 10^{24}$
cm$^{-2}$; absorption of mid-IR lines, molecular maps, and the great
difference in the average X-ray columns between Type 1 and Type 2 AGN suggest
that a commensurate dust extinction is present. \citetext{See 
\citealp{2001A&A...365...28M,2001A&A...365...37M,2001A&A...375...25M}
for arguments which affect this line of reasoning quantitatively but not 
qualitatively.}

There is a limit on the anisotropy of the ratio [O III] $\lambda$5007/F$_\nu(60\mu)$
in Seyferts from Fig.~3 of \citet{1994A&A...283..791K}. The figure shows that in their
well-selected sample ($60\mu$ flux with a mild $25\mu$--$60\mu$ warmth criterion),
Type 1 and Type 2 objects have indistinguishable distributions of L($60\mu$),
L[$\lambda$5007], and of course their ratio. The $\lambda$5007 line is produced outside
the torus in most objects like these: it doesn't appear in polarized flux along
with the broad emission lines and Big Blue Bump. Thus it's not significantly
hidden inside the torus, is likely quite isotropic in this parameter space, and
is nearly isotropically selected\footnote{\setlength\baselineskip{1ex}
Isotropic selection means selection on
an isotropic AGN property, which avoids powerful biases in comparisons between
classes. It is essential in order to produce intelligible results.}, fairly powerful Seyferts.

Torus models do predict that the optical depths will be small in the far-infrared
in general. For AGN-dominated infrared SEDs, that means there is an elegant
and detailed method of deriving the degree and wavelength-dependence
of the dust emission anisotropy. For isotropically selected samples, we can
divide composite or representative Type 1 SEDs by those for Type 2, tying them
together at $60\mu$.

In order to make the infrared calorimeter accurate, we need to account for the
common anisotropy of the near- and mid-thermal dust emission. There is general
agreement on near isotropy past $\sim30\mu$, as was first predicted by
\citet{1992ApJ...401...99P,1993ApJ...418..673P}. The main basis for their
prediction of anisotropy at shorter wavelengths was that many Seyfert 2's are
Compton-thick, including NGC1068 --- which is completely opaque and for which
$N(H)\gtwid10^{25}$ according to the X-ray spectrum \citetext{e.g., \citealp{2006MNRAS.368..707P}}.
In Galactic dusty gas this corresponds to $A(V) \sim 1000$. \citetext{See e.g.,
\citealp{2001A&A...365...28M,2001A&A...365...37M} for a
quantitative correction, which however doesn't greatly
affect the discussion of gas vs.\ dust columns below.} As an aside, I do think
that dust-free atomic gas can contribute to the X-ray absorption in some cases
\citetext{e.g., \citealp{2011MNRAS.410.1027R,2004ASPC..311..381A}}, but the X-ray columns of
Type 2 AGN are on average $\gtwid100\times$ those of Type 1, so most of the column is
connected with the Type 2 classification. Also in some cases, the dust column
can be constrained to be similar to the very high X-ray columns. For example,
\citet{2000ApJ...530..733L} used the lack of Pf-$\alpha$ at $7.46\mu$ to show that
$A(V) > 50$ in NGC 1068.

Recall that one could pick spectral Type 1 (broad-line) radio loud quasars and
radio galaxies by some fairly isotropic AGN-related luminosity (e.g.,
lobe-power), and divide the two spectra to produce a spectrum of diminuation
from anisotropy. We were able to do this for the $z>1$ 3CR (Fig.~4) because all
of the radio galaxies seem to host hidden quasars, but only out to $\sim15\mu$
in the rest frame. {\it The correction for anisotropy is around a factor
of 10 in the near-IR, 1.5--2 on both sides of the silicate feature,
and about three in the silicate feature}, but varies significantly from object
to object. As expected, it flattens out at $\sim1$ at long wavelengths.

\begin{minipage}{5in}
\vskip 4truein
\includegraphics{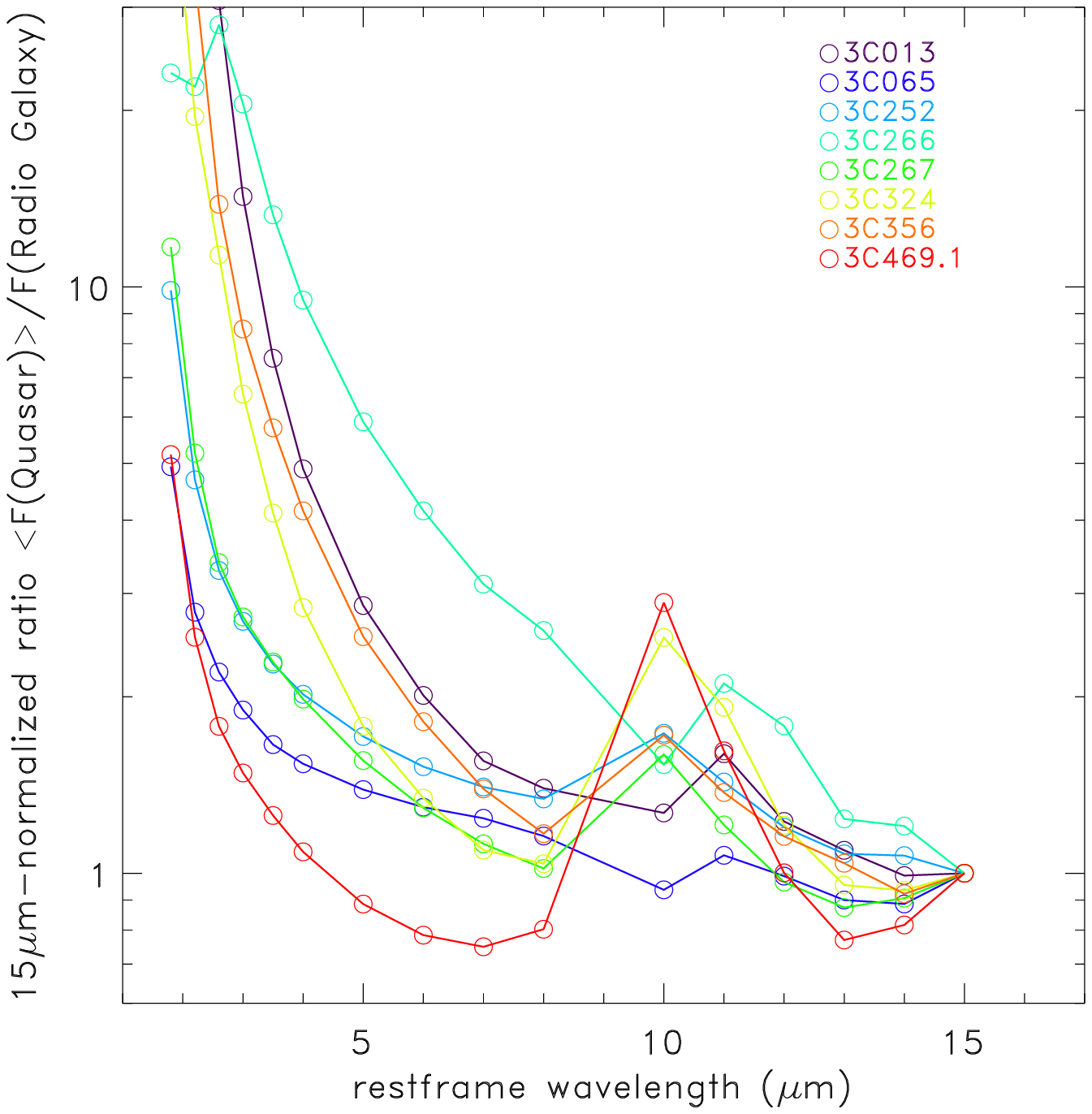}
\setlength{\baselineskip}{1ex}
\vskip 4mm
{Figure~4. The infrared SEDs of $z>1$ 3CR quasars and the matched radio galaxy
composite, from \protect\citet{2011ApJ...736...26H}.
Since these radio galaxies are all edge-on quasars, this quotient spectrum
measures the orientation dependence at each wavelength.}
\end{minipage}
\vskip 2.5mm

Recall that torus shape was inferred\footnote{\setlength\baselineskip{1ex}
A common error is to use the
percent polarization of the continuum as the scattering polarization (after
correcting for bulge or elliptical starlight). That is usually contaminated by
hot stars, and only the broad lines themselves can be used to find, or more
often place, a \underbar{lower limit} on, the percent polarization of the scattered
nuclear light \protect\citep{2002ASPC..284..147A}.} from the high typical polarization of the
reflected light in hidden-Broad Line Region objects, which is generally perpendicular
to the radio axis, meaning that photons can only escape the nucleus to scatter
into the line of sight if they leave the nucleus in the polar directions.

The obscuring dusty tori invoked in the shadowing aspect of the unified model
are active, not just passive, components. Modulo factors of order unity for
geometry and dust cloud albedo, the tori will reprocess almost all of the
incident Big Blue Bump/Broad Line Region luminosity into the infrared. Thus the
ratio of re-emitted infrared emission to Big Blue Bump (and often considerable
absorbed X-ray) emission tells us the approximate covering factor of the dusty
gas which is idealized as a torus shape.  There is at
least one obscured quasar with a \underbar{thin} obscuring disk --- the
radio-loud, mini-MgII BAL OI287 \citep{1988ApJ...331..332G,
1988ApJ...331..325R,1988ApJ...328..569U,1993ApJ...414..506A}; however, even
that one seems to have an upturn longward of $1\mu$, at least according to the
NED figures.

The covering fraction should agree with the fraction of Type 2 (hidden) nuclei
in a sample which is selected by any isotropic AGN property. What are the
results of this comparison? I'm crudely leaving out intermediate types, and the
substantial minority of Seyfert 2s whose Type 1 nuclei are hidden by dust in the
host galaxy plane \citep{1980AJ.....85..198K,1982ApJ...256..410L} rather than by a nuclear
torus. \citet{1994A&A...283..791K} found 80 far-IR selected Seyfert 1s and 141
Seyfert 2s (plus some H II galaxies that they weeded out).
So one expects a dust covering factor of $\sim$ two-thirds in this parameter space.
While the paper describes dust and gas covering factors as ``broadly consistent''
with the model, it would still be very valuable at this point to take advantage
of this sample and actually measure covering factor for the Type 1s based on
existing spectral energy distributions.

Figure~4 \citetext{from \citealp{2011ApJ...736...26H}} shows the quotient SED for $z>1$
individual quasars and the radio galaxy composite, which however only covers up
to 15 microns in the rest frame.\footnote{\setlength\baselineskip{1ex}
\protect\citet{2006AJ....132..401B}
does this type of division for a $12\mu$ (better than optical, but not ideal)
selected sample of Seyfert galaxies. As expected, their anisotropy curve
flattens at long wavelengths, but oddly not at a value of unity.} Again this
plot purports to give us directly the anisotropy as a function of wavelength.

\section{CLASSICAL DOUBLE, OR FR II RADIO SOURCES}
\subsection{Spectropolarimetry of Radio Galaxies and the Discovery of Hidden Quasars}
Hidden quasars inside radio galaxies are discussed in \citet{1993ARA&A..31..473A,
1999ASPC..162..101W}, and \citet{2008NewAR..52..227T}. I'll describe many relevant
observations here, moving from the radio and then on up in frequency to the X-ray.
First though, I will introduce some background optical information, and will
return to that waveband in Section~\ref{sec:classic-xray}.

It was noticed in the early 1980s that for some Narrow Line Radio Galaxies and
Seyfert 2 galaxies, a small measured optical polarization could often be
intrinsically large ($\gtwid10$\%) after starlight subtraction, and that the
electric vector position angle was generally perpendicular to the radio
structure axis \citep{1982Natur.299..605A,1983Natur.303..158A,2002apsp.conf..151A,
1983ApJ...271L...7M,1983Natur.304..609M,1993MNRAS.262.1029D}.
Furthermore, the polarized light spectrum (similar to the scattered
light spectrum in spectral features) revealed the features of Type 1 AGN, with
the first case being the powerful hidden quasar in the radio galaxy 3CR234 shown
in Fig.~2 \citep{1982Natur.299..605A,1984ApJ...278..499A,1995AJ....110.2597T,
1998MNRAS.294..478Y}. As noted earlier, this means that the galaxies contain
hidden quasars, and that their photons can reach us by exiting the nuclei along
the (radio) structural axes and then scattering into the line of sight. The
\citet{1995AJ....110.2597T} polarized flux spectrum is shown in Fig.~2.

Since then many other radio galaxies, in general those with strong high-ionization
lines like 3CR234, have shown hidden quasars in polarized light. These papers
include most of them: \citet{1995AJ....110.2597T,1998ApJ...500..660T,1999AJ....118.1963C,1996MNRAS.279L..72Y,
1996ApJ...465..157D,1996ApJ...465..145C,1997ApJ...476..677C,1997ApJ...482L..37O}
on Cygnus A \citetext{see also \citealp{1994Natur.371..313A}}; \citet{1999AJ....118.1963C,1999ApJ...514..579H,
2001ApJ...547..667K,2001A&A...366....7V,2002MNRAS.330..977T,2004MNRAS.351..997S,2005ASPC..343..457T}.

Many Seyfert 2s were subsequently shown to have hidden Type 1 nuclei by this
method \citetext{\citealp{1985ApJ...297..621A}; many references are given in
\citealp{2003ApJ...583..632T}}, but radio loud cases were fewer, at least
partially because they are more distant and fainter.

While there is no question that 3CR234 hosts a powerful hidden quasar, it has
some properties which make it somewhat special as a radio galaxy: the luminous
high-ionization narrow lines and the powerful infrared dust source \citetext{see Fig.~3;
also \citealp{1998MNRAS.294..478Y}}. Most of the other radio galaxies shown to
host hidden quasars in this way share these properties \citep{1999AJ....118.1963C}.
Both properties are rare in FR I radio galaxies, and at low and moderate redshift,
many FR IIs differ from this pattern as well \citetext{e.g., \citealp{1979MNRAS.188..111H};
Table 2 of \citealp{1981ApJ...243...81C} on optical spectra; \citealp{2006ApJ...647..161O},
\citealp{2009ApJ...694..268D} on infrared observations}. Many of these have
low-ionization, low luminosity emission lines \citep{1979PASP...91..257M}.
(All of my reference lists are undoubtedly incomplete!)

Radio galaxies with $z\gtwid1$ usually have resolved optical light, and much of
the light is scattered from hidden quasars \citep{1987Natur.329..604C,1987ApJ...321L..29M,
1993MNRAS.263..936D,1993MNRAS.264..421C}. Spectacular exceptions include 4C41.17
\citep{1997ApJ...490..698D} and 6C 1908+722 \citep{1999mdrg.conf...19D}; they
show optical light extended along the radio axes, but it is mostly unpolarized
starlight. However, these $z\sim4$ objects are observed at higher rest-frame
frequencies than explored in other objects.

\subsection{Puzzling statistics on the radio properties of FR II radio galaxies and quasars}
There are a few historical papers which are interesting and illustrative of the
reasoning that led to much progress on unification. Peter Barthel worked
mostly on VLBI observations of superluminal sources in the 1980s, noting that
the beam model explained many properties such as superluminal motions and jet
sidedness qualitatively, but he was (according to the title of a rumination for
a conference) ``feeling uncomfortable'' because one had to assume that a large
fraction of these sources in various samples have jet axes fortuitously close to
the line of sight.\footnote{\setlength\baselineskip{1ex}
When selecting by high-frequency radio flux, one
preferentially selects objects whose jet axes point in our general direction
because of beaming, but this effect can't explain the apparent preference for
this orientation quantitatively.} Barthel later wrote a famous paper
\citep{1989ApJ...336..606B} entitled ``Are all quasars beamed?'' suggesting that
those quasars whose axes lie near the sky plane
somehow fall out of quasar samples, and (inspired by the spectropolarimetry)
might be classified as radio galaxies. Note though that he did not entitle his
paper ``Do all radio galaxies contain hidden quasars?'' The answer to that
question would be no, but the answer to the question he posed is still basically
yes.

The general idea of beaming to explain superluminal motions and one-sided jets
was accepted by most doubters as a result of two key discovery papers reporting
on the so-called
lobe depolarization asymmetry. There is a very strong tendency for one radio lobe
in double-lobed radio quasars to be depolarized at low frequencies by Faraday
rotation within the observing beam on the side of the single-jet sources which
{\it lacks} the jet \citep{1988Natur.331..149L,1988Natur.331..147G}. Most people
accepted that the depolarized lobe must be the more distant one, located behind
a large-scale depolarizing magnetoionic medium; thus the polarized lobe is on
the near side, so that the jet is also on the near side, as expected for
beaming. \citetext{A demur can be found in \citealp{1989AJ.....97..647P}.}

The selection criteria in \citet{1988Natur.331..149L} and in \citet{1988Natur.331..147G} favored
low-inclination sources; nevertheless it is amusing that the former paper has
this disclaimer: ``The sources observed here must then be oriented within about
45 degrees of the line of sight\dots to generate sufficient asymmetry in path
length\dots'' to fit the depolarization data. Their referees must not have been
particularly curious people not to ask for elaboration. We now know that many
of the high-inclination sources were masquerading as FR II radio galaxies.

\citet{1989ApJ...336..606B} focused on the 3CR sources in the redshift interval $0.5 < z <
1.0$, in order to avoid low-luminosity, especially FR I radio galaxies, which he
did not propose to identify with quasars, noting that ``radio loud quasars
invariably\footnote{\setlength\baselineskip{1ex}
``Almost invariably'' would have been more accurate, e.g.,
quasar 1028+313 in \protect\citet{1984AJ.....89.1658G}} have (lobe) luminosities
in excess of the Fanaroff and Riley division.'' He also pointed out that the 3CR
radio galaxies above $z = 0.5$ have strong emission lines like
quasars.\footnote{\setlength\baselineskip{1ex}
Today we would say that the large majority of them have strong
emission lines.} He notes further that, based on early fragmentary data, only a
subset of FR II radio galaxies are strong IR dust emitters. Finally, Barthel
crucially adds this claim later in the paper: ``\dots including the $0.3 < z < 0.5$
and/or the $1.0 < z < 1.5$ redshift range does not alter [his conclusions] markedly.''

I was particularly moved a few years later by a paper by \citet{1993MNRAS.262L..27S},
entitled ``Evidence against the Unified Scheme for Powerful Radio Galaxies and
Quasars.'' I reproduce two of his figures here (the present Fig.~5 and 6).
His histograms of number densities and cumulative linear size distributions are
similar to Barthel's for $z > 0.5$, but he includes low-redshift FR II sources
in a new $z < 0.5$ bin, where ``all hell breaks loose.'' The median projected
linear size in quasars relative to radio galaxies no longer shows a satisfying
reduction expected for foreshortening, and the number density of FR II radio
galaxies becomes much higher than those of quasars.

Singal suggested that this spoiled the Unified Scheme, but a clever alternative
was suggested by \citet{1996ApJ...463L...1G}. They showed that subject to two
assumptions justified or at least motivated by independent observations,
Singal's histograms could all be easily understood in the beam model. One
hypothesis was that the torus opening is set by the initial radio power of a
source; the other was that the luminosity of a growing giant double radio source
decreases over time in a certain way. Without going through all the reasoning,
it turns out in this case that in the lowest redshift bin, one is preferentially
comparing older quasars to younger radio galaxies, canceling (to the modest
accuracy attainable) the expected size difference between the two types of radio 
source.

There is another possible explanation for Singal's histograms: there is a
population of FR II radio galaxies, concentrated at low redshift, which simply
lacks hidden quasars.\footnote{\setlength{\baselineskip}{1ex}
As we will see, some FR II radio sources do lack
quasars, even hidden quasars, and they should form a roughly isotropic
distribution. So an easy study would be to look at the strengths of their
depolarizations. Another easy armchair ApJ Lett would be to re-create the
Singal plots, but leaving out the nonthermal galaxies (no hidden quasars), which
should be fairly isotropic and whose removal should make the $z < 0.5$ bin look
more like the other bins.}

\begin{minipage}[b]{5in}
\vskip 4truein
\begin{minipage}{2.5in}
\includegraphics{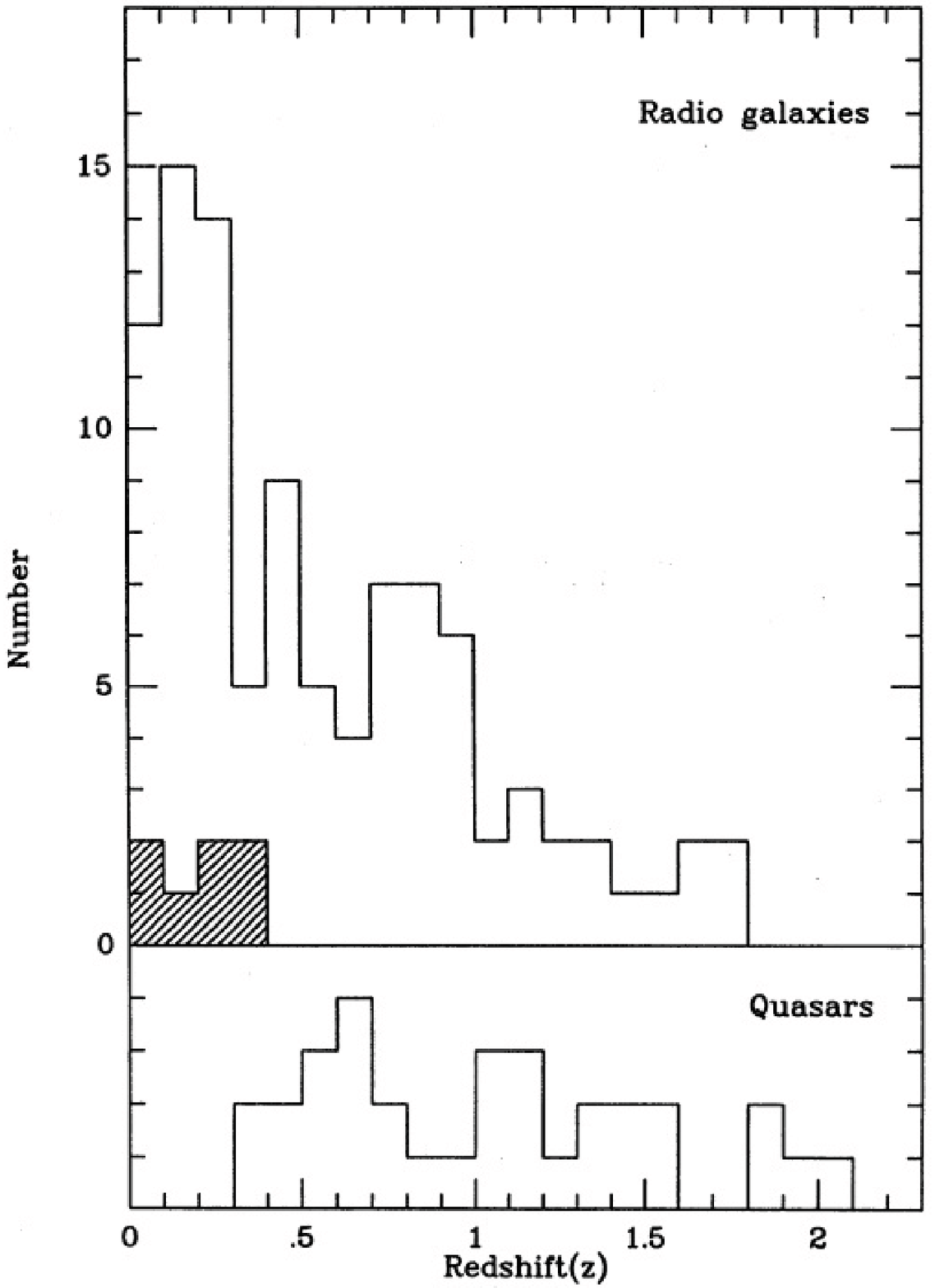}
\setlength{\baselineskip}{1ex}
{Figure 5. Redshift distribution of 131 radio galaxies and quasars in the 3CR
sample, taken from \citet{1993MNRAS.262L..27S}. Crucially, the hashed ``Broad
Line Radio Galaxies'' must be mentally moved to the lower plot for the purposes
of this paper. Note the large ratio of the number of radio galaxies to that of
quasars at low redshift.}
\end{minipage}
\hspace{0.25truein}
\begin{minipage}{2.5in}
\includegraphics{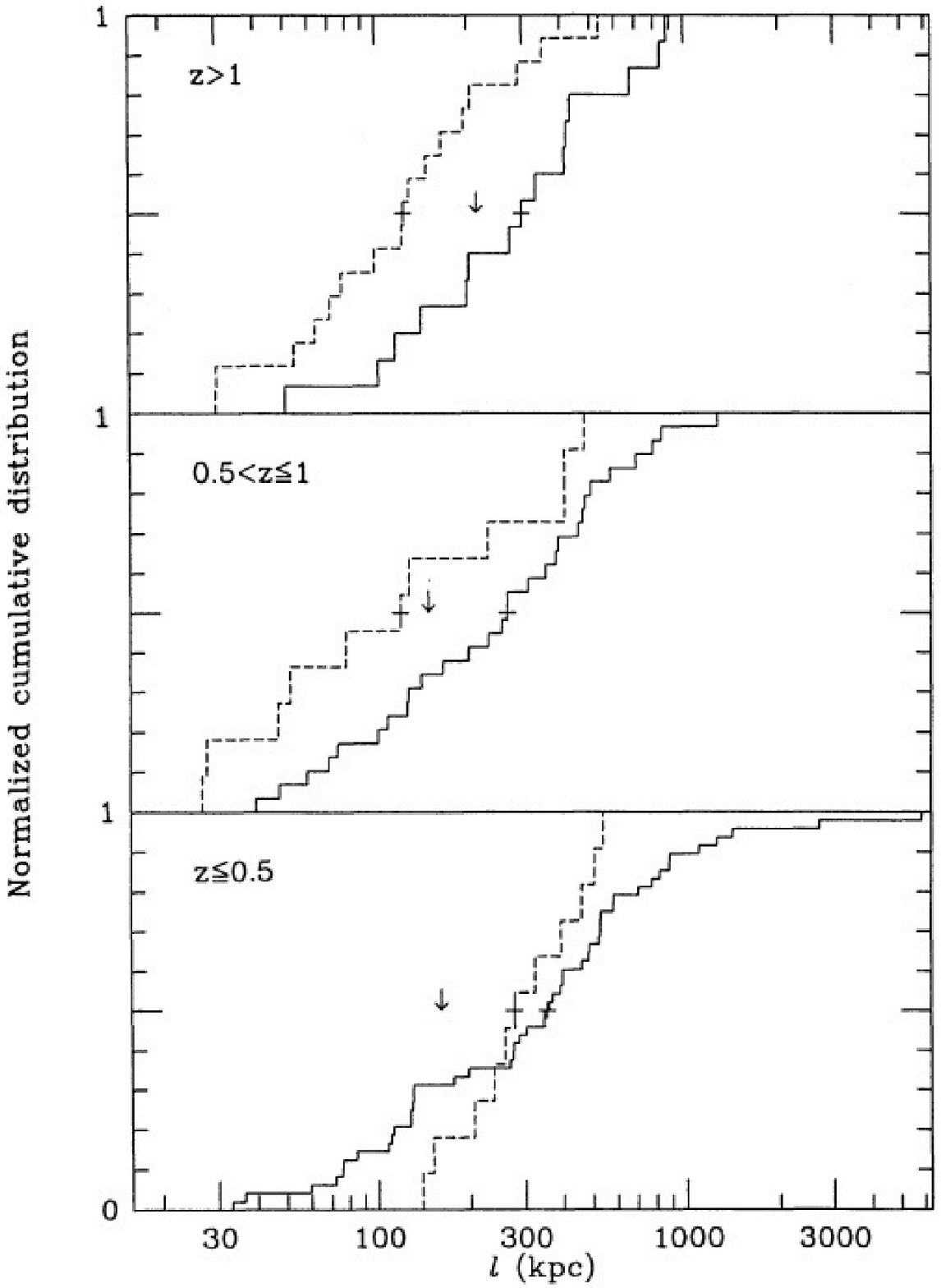}
\setlength{\baselineskip}{1ex}
\vskip 4mm
{Figure 6. Cumulative distributions of projected linear sizes of radio galaxies
(continuous curve) and quasars (dashed curve) in several redshift bins.
Crosses mark the median values, and arrows are for a test used by the
author, but not referred to here \protect\citep{1993MNRAS.262L..27S}. The
foreshortening expected in the shadowing Unified Model is not seen at low redshift.}
\end{minipage}
\end{minipage}
\vskip 2.5mm

\subsection{The infrared calorimeter for visible and hidden FR II radio sources}
Distinguishing these two hypotheses motivated my group and others to pursue
observations of radio galaxies in the thermal infrared, where the infrared
calorimeter (radiation reprocessed as infrared) must show us the putative hidden
quasars, independently of orientation. Remember that the spectropolarimetric
test for hidden AGN requires a somewhat fortuitous geometry, where a gas and
dust cloud must have sufficient optical depth and covering factor to reflect
detectable light, and it must have a view of both the hidden quasar and the
observer on Earth. Thus it is an incomplete method for finding hidden AGN.

David Whysong and I started imaging 3CR radio galaxies and quasars from
Singal's list at Keck Observatory in the 1990s, but equipment problems,
terrible weather, and refractory referees delayed our project until
ISO and even Spitzer were making big radio galaxy surveys, albeit at low
angular resolution. The power and elegance of the infrared calorimeter
is shown best however in the composite of high-resolution ($\sim0.3^{\prime\prime}$)
images reproduced from Whysong's (\citeyear{2005PhDT.........4W}) thesis, and
shown here as Fig.~7.

\begin{minipage}{5in}
\vskip 3.5truein
\includegraphics{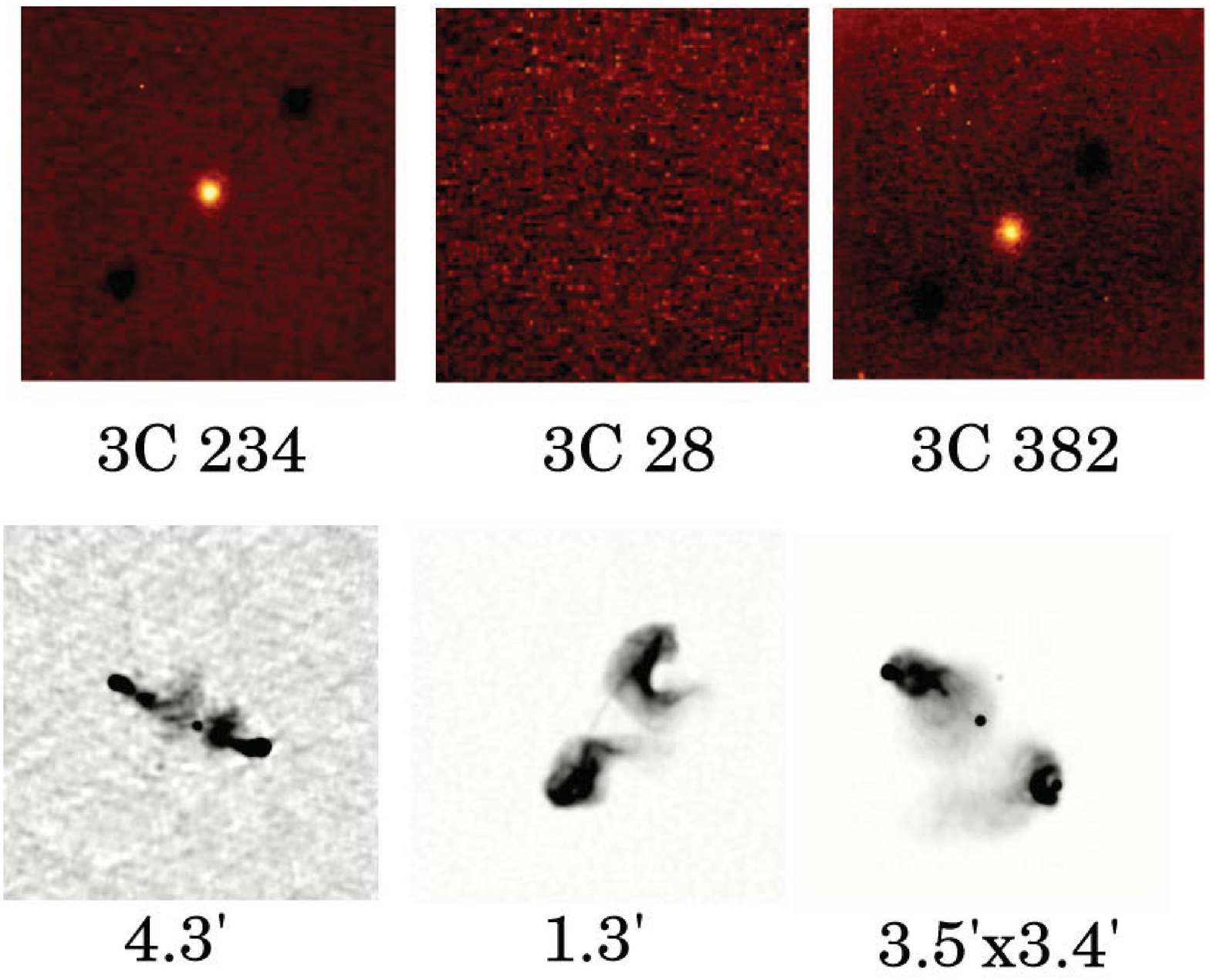}
\setlength{\baselineskip}{1ex}
\vskip 4mm
{Figure 7. Keck images of three radio-loud FR II AGN, taken in the thermal IR at
11.7$\mu$. 3CR382 is a visible Broad Line Radio Galaxy. 3CR234 is known to be
a hidden quasar from its optical polarized flux spectrum. Apparently,
3CR28 does not hide a powerful quasar since no infrared reradiation is
detected. (Note that the apparent correlation of radio core strength with
central engine type is not a general property.) From \protect\citet{2005PhDT.........4W}.}
\end{minipage}
\vskip 2.5mm

\subsection{Infrared properties of FR II quasars and radio galaxies from 
the 3CR catalog\label{sec:3.4}}
Many groups have been seduced by the seeming simplicity and robustness of the
infrared calorimeter. A rather complete review of early (pre-Spitzer)
observations is given in the Introduction to \citet{2007ApJ...660..117C}, reporting on
Spitzer observations of 3CR objects with $0.5 < z < 1.4$. Other recent papers
include \citet{2001A&A...372..719M,2004A&A...421..129S}; and \citet{2004ApJ...602..116W},
including some $0.3^{\prime\prime}$-resolution Keck data at $11.7\mu$.

\citet{2001A&A...372..719M} observed 20 3CR objects with ISO, finding the
results compatible with hidden quasars, {\it except possibly for some at the
low luminosity/redshift end}. Relatedly, \citet{2004A&A...421..129S} found 68
detected 3CR objects in the ISO archive, finding that, ``In most 3CR objects,
the mid- and far-IR flux cannot arise from stars nor from the radio core
because an extrapolation of either component to the infrared fails by orders of magnitude.''

\citet{2005ApJ...629...88S} used Spitzer photometry to study a sample of 3CR
radio galaxies and steep-spectrum quasars with good Hubble Space Telescope
(HST) images, and with ``a preference for $z < 0.4$''. The data are consistent
with the assertion that {\it most radio galaxies have hidden quasars}.
Furthermore, there is evidence for nonzero optical depths at $24\mu$.
Importantly, the entire infrared dust spectrum is generally inferred to be AGN
powered, even at $70\mu$.

Spectroscopy in the optical and the infrared can deliver more specific
information. Sometimes the optical narrow line spectrum indicates a lower
ionization or luminosity than is actually present, and thus one could conclude
that a hidden quasar is not present. Again infrared photometry and spectroscopy
are more complete. \citet{2005A&A...442L..39H} wrote a key paper which must be
kept in mind whenever narrow line spectra are discussed. I hope more work is
done along these lines. \citet{2005A&A...442L..39H} found that in a set of
seven radio galaxies and seven quasars from the 3CR catalog, the radio galaxies
observed in the optical have on average less luminous high ionization lines
than the quasars, but that in the relatively transparent mid-IR region, for
this small sample of FR IIs at least, {\it this is not the case}.
Haas et al find that ``the luminosity ratio
[OIII]$\lambda$5007\AA\ / [OIV]$25.9\mu$ of {\it most} galaxies is lower by
{\it a factor of 10} than that of the quasars''!\footnote{\setlength\baselineskip{1ex}
Note that this large
anisotropy or at least absorption of [O III]$\lambda$5007 differs markedly from
the situation in the Seyfert sample discussed in Sec.~\protect\ref{sec:broadinfraredcal}.}

Similarly, \citet{1997A&A...328..510D} showed that for some powerful radio
galaxies at least, [O III]$\lambda$5007\AA\ shows up to some degree in polarized
flux, indicating some [O III]$\lambda$5007\AA\ is probably hidden by the torus
in some cases; but note that it's not always easy to interpret the small
polarizations of [O III]$\lambda$5007.

Another important consideration is that at least part of the obscuration of the
Narrow Line Region in radio galaxies is due to the modest optical-depth
kpc-scale dust lanes as seen so dramatically in Cyg A and Cen A. See discussion
in \citet{1990ApJ...363L..17A}. Ignoring this information in testing and
exploiting the Unified Model will produce erroneous results. Absorption by
cold foreground dust is manifest as very deep mid-IR absorption features.

Now let's go back to the figures from \citet{1993MNRAS.262L..27S}, here Fig.~5
and 6. We noted that at low redshift, a {\it large fraction} of 3CR FR II radio
galaxies are {\it smaller than expected} for high inclination quasars. This
could be interpreted as evidence that there are many intrinsically small radio
galaxies which lack powerful hidden quasars. By powerful, I mean hidden quasars
roughly matched in reradiated Big Blue Bump luminosity with radio galaxies of
the same lobe flux and redshift. (Some allowance needs to be made for mid-IR
anisotropy.) Alternatively, recall that \citet{1996ApJ...463L...1G} cleverly
noted that two externally motivated assumptions would cause a universally
applicable unified model (hidden quasars in all radio galaxies) to lead to
data matching the observations; one has to assume that radio sources
tend to fade over time, and that the opening angle of the torus increases with
the {\it original} radio power. Our group was fortunate to receive time
for Spitzer IRS surveys of several types of radio-loud AGN, including 3CR FR II
radio galaxies and quasars \citep{2006ApJ...647..161O,2007ApJ...668..699O}.
The 2006 paper was the first from a big Spitzer IRS-spectrograph survey.

These infrared data show in fact that many of the 3CR FR II radio galaxies,
especially those at $z\ltwid0.5$ \citetext{below the redshift considered in
\citealp{1989ApJ...336..606B}},
have only weak and cool dust emission. Again if many of these objects have
AGN hidden by dust, then the average dust covering factor would need to be reasonably
large, so that the dust calorimeter should work at least statistically. Most 3CR
FR II radio galaxies with $0.5 < z < 1.0$ do have strong, quasar-like mid-IR
emission. This is qualitatively in accord with both Barthel's and Singal's figures
(which suggests a dearth of hidden AGN in FR IIs only at $z<0.5$). Also we now
know from Spitzer that essentially {\it all} the $z \gtwid1$ radio galaxies in
the 3CR can be unified with quasars by orientation \citetext{\citealp{2006ApJ...647..161O,
2008ApJ...688..122H,2010ApJ...717..766L,2010ApJ...725...36D}, which covers
(non-3CR) galaxies up to $z = 5.2$}. Compare Fig.~8, showing mid infrared
luminosity vs.~$z$, with Figs.~5 and 6 \citetext{from
\citealp{1993MNRAS.262L..27S}}.\footnote{\setlength\baselineskip{1ex}
One can also see this effect to some extent
in the closely related top parts of Fig.~8 of \protect\citet{2007ApJ...660..117C}.}

Now let's see how the accretion luminosities, taking account of the IR data,
compare with $L_{\rm Edd}$ (Fig.~9). Among these radio galaxy black holes, the
ones considered to be hidden quasars (for which probably $M_{BH}\sim10^9 M_\odot$)
radiate at $\sim0.1$\% of Eddington in dust reradiation only, and the others
are below that value. (Figure~9 is a preliminary version of this plot, with some
subjective elements, kindly supplied by P.~Ogle 2010.) Although this cut isn't always
accepted by nature, it's interestingly close to the value expected for the shift
from ADAF to thermal optically thick Big Blue Bump \citetext{$\dot m_{\rm crit}=1.3\alpha^2$
which produces $L\sim0.13$\% of $L_{\rm Edd}$, for 10\% efficiency in $\alpha=0.1$ 
models: \citealp{1997ApJ...489..865E}}.

\vskip 5mm
\begin{minipage}{5in}
\vskip 2.75truein
\includegraphics{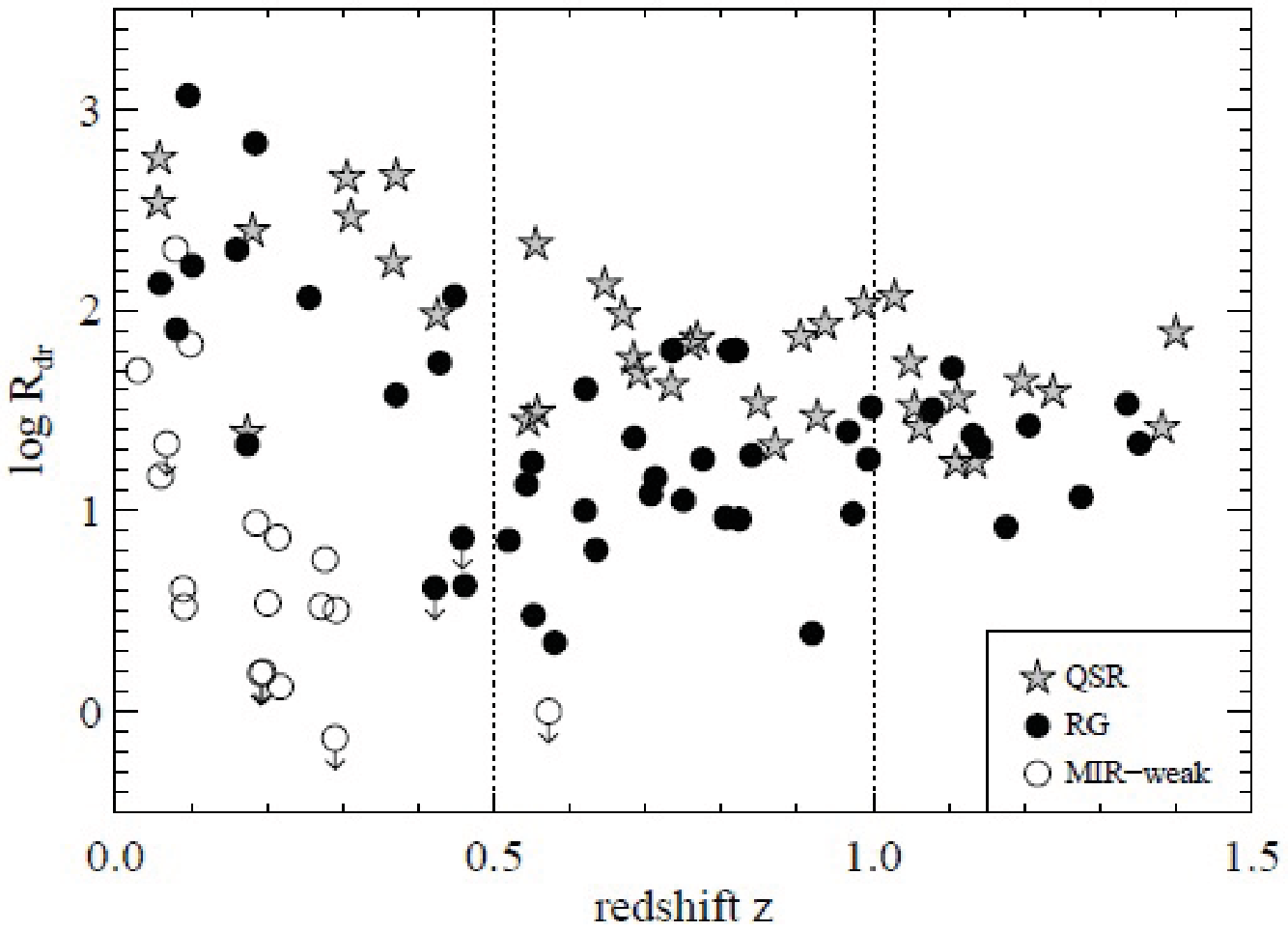}
\setlength{\baselineskip}{1ex}
\vskip 4mm
{Figure 8. The ordinate shows the 15 micron luminosity $\nu L_\nu$, normalized
to the rest-frame 178 MHz luminosity, also referring to $\nu L_\nu$. Plotting this
normalized infrared luminosity allows us to compare objects of fixed power in the
nearly isotropic diffuse radio emission. Thus, there are no selection biases with
respect to orientation and so the objects in various categories are directly comparable.
Stars represent quasars and filled circles refer to the putative hidden quasars,
with L(MIR)$\sim$L(MIR) for visible quasars. Empty circles lack visible or hidden
quasars, and their presence at low $z$ can explain the present Figs.~5--7. (see text).}
\end{minipage}
\vskip 5mm

If one were to display the apparent {\it optical} luminosities in
Eddington terms, there would be no change in narrow line ionization (see $H$
and $L$ symbols) at this expected location in $L/L_{\rm Edd}$ of the 
accretion mode change. But using the infrared
luminosities\footnote{\setlength\baselineskip{1ex}
\protect\citet{2004MNRAS.351..733M} present
a proxy for this test which could be made without IR data. Using line emission
to estimate bolometric luminosities for radio galaxies, they found a probable
gap in $L/L_{\rm Edd}$ at $\sim0.01$.} seems to confirm theory spectacularly,
if approximately. It is also very clear that both the ionization level and
(more surprisingly) the FR type are tightly correlated
with accretion mode, as deduced by many authors over the years, and cited in
context in this paper.

When a galaxy lacks an observable Big Blue Bump, even when using the IR reradiation to
find it, we can of course only put an upper limit on the flux of that component.
According to theory, objects with $L/L_{\rm Edd}\ltwid0.01$ aren't expected to
produce optically thick accretion disks (Big Blue Bumps), so perhaps few weak
ones are missed \citep{1984RvMP...56..255B}. However, the theory
isn't yet robust enough to be used in this way with certainty.

There are several fiducial luminosity comparisons that we can make to our infrared upper
limits in the radio galaxies: we can compare them to those of matched broad line objects
to constrain the universality of the unified model. More important physically,
we can compare it to the jet power. The latter is hard to get, and other than
for the nearby and powerful object M87, the infrared limits are too high 
to be of interest in comparing to jet
power. But for M87, any hidden quasar must produce much less radiative than
kinetic luminosity \citep{2004ApJ...602..116W,2000ApJ...543..611O,2001ApJ...561L..51P}.

\vskip 5mm
\begin{minipage}{5in}
\vskip 3.75truein
\includegraphics{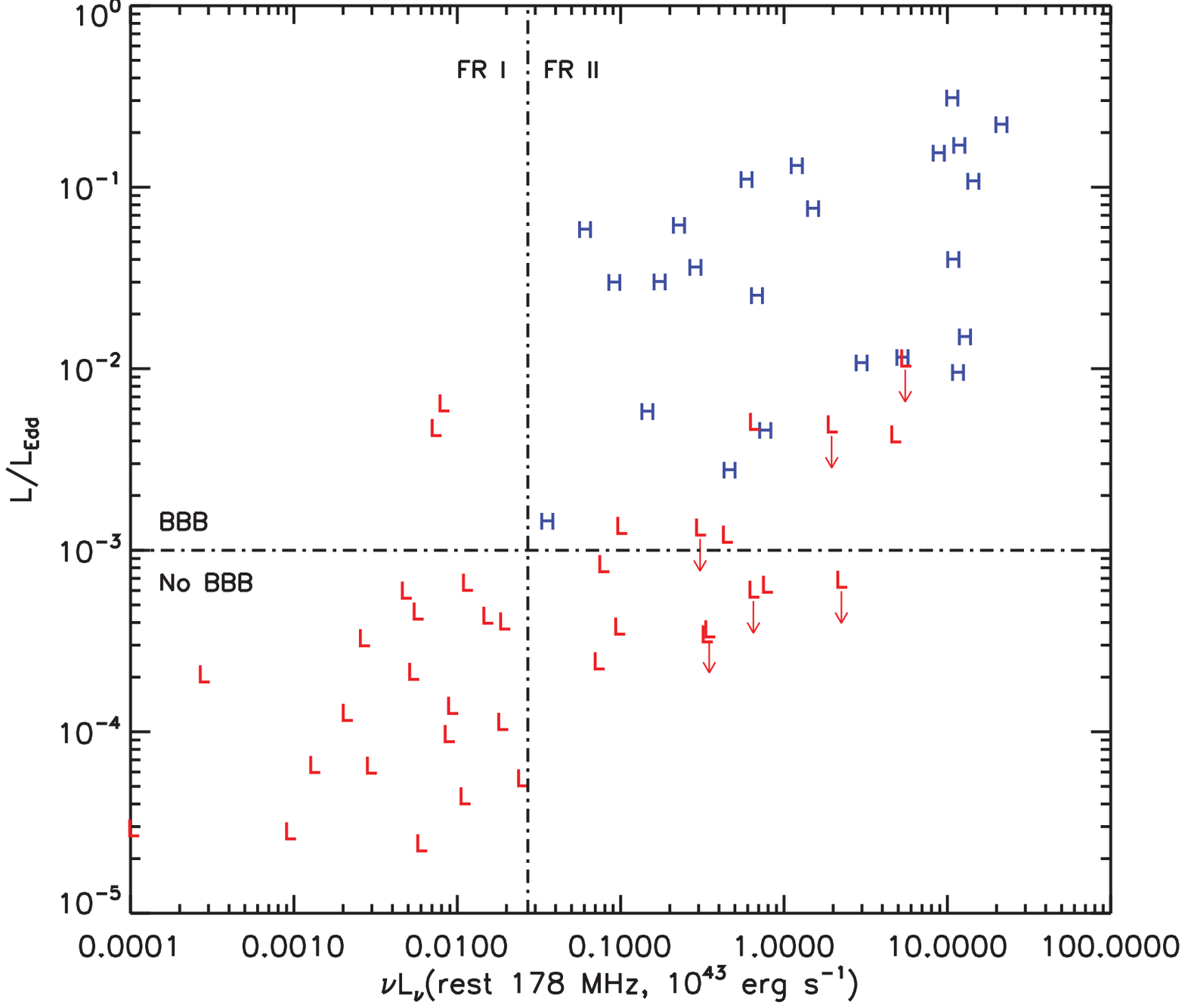}
\setlength{\baselineskip}{1ex}
{Figure 9. This plot of 3CRR radio sources is preliminary and involves some
judgment regarding covering factor, absorption inside the torus, infrared
anisotropy, and bolometric corrections. This plot, and the details about these
assumptions, can be found in \protect\citet{2011ApJS..195...19O}. Masses are
based on spiral bulge or elliptical host luminosity. Uncertainties are a factor
of several. Note that the objects with quasar-like mid-infrared luminosities are
almost all of type FR II, as well as high in ionization level. Details in
\protect\citet{2010ApJ...724.1193O}}
\end{minipage}
\vskip 5mm

I end this section with some general caveats on unified models and accretion
modes for FR II radio sources. 1) We're discussing the
highest-luminosity radio sources at each redshift, and in the case of the
complete geometrical unification at $z > 1$, we're only talking about some of
the most luminous radio sources in the universe. 2) The statistical
significance of the Singal figures makes them robust, but it isn't sufficient
for any further investigation by subdivision. 3) At least at the highest
redshifts, we know that there is a major contribution to the projected linear
sizes of radio sources besides foreshortening. It was shown by \citet{2000MNRAS.311...23B}
and subsequent papers that the $z\sim 1$ 3CR radio galaxies' projected linear
sizes in the extended gas correlates strongly with ionization level, the smaller
ones tending to have shock spectra. Aside from adding noise to the radio size
tests, this doesn't affect the discussion too much, though of course it's
important from a physical point of view. Note also that that correlation is shown
in \citet{2000MNRAS.311...23B} for
the radio galaxies alone, where orientation is relatively unimportant, so the
claim would have to be softened if made for the entire high-$z$ 3CR sample.

\subsection{Nonthermal optical ``compact cores''}
Several authors have commented on the optical point sources seen even in some
Narrow Line Radio Galaxies which lack a visible Type 1 spectrum.  This is different
behavior than that in Seyfert 2s, few of which have point sources or variability ---
their mirrors are extended and spatially resolved by the Hubble Space Telescope
in nearby cases, e.g., \citet{1995ApJ...446..155C,1995ApJ...452L..87C};
Kishimoto \citeyear{1999ApJ...518..676K,2002ApJ...565..155K,2002ApJ...567..790K}.

It's artificial to separate the FR Is from the FR IIs in this context, because
the entire radio galaxy population empirically separates itself in a different
way: the large majority of 3CR FR Is (these are nearby, $z \ltwid0.2$), and
many low ($z < 0.5$) and some intermediate ($0.5<z<1.0$) redshift
3CR sources have optical nuclei which are consistent with emission from the
unresolved bases of the radio jets. References are given below.

In general there are no spectra available of these optical point sources, and
certainly no spectropolarimetry. However, the red-region HST point source
luminosities {\it and fluxes} correlate fairly well with the 5GHz (usually
flat spectrum) radio cores, which are indeed the bases of jets as shown by VLBI
maps. It's important that the correlation shows up in a flux-flux plot as
well as in a luminosity-luminosity plot. In luminosity-luminosity plots of AGN (radio
loudness vs.\ optical power is an example of an exception), you will almost
always see a correlation because more powerful objects tend to have more of
everything. Flux-flux plots have their own peculiarities, but they are {\it different}
peculiarities than in luminosity-luminosity.

Two other very common statistical errors that drive
me crazy are: 1. in plots of the form A vs.\ A/B (or B/A), which are very common, the
correlation slope, which may be intrinsically zero, will be strongly biased
towards a positive (negative) value --- it is {\it not} a discovery when this
happens, if the ordinate range due to errors and population dispersions isn't
{\it much} smaller than the range of the ``correlation.'' People actually
publish plots like this all the time, then go as far as analyzing these spurious
slopes and trying to extract physics from them. 2. Survival statistics are often
used to deal with upper limits, but most astronomical data sets violate
the key requirement for this method: that the limits have the same distribution
as the detections. In astronomy, we tend to have the exact opposite case: that
the upper limits are usually concentrated towards the bottom of the distribution
of detections!

Chiaberge and collaborators have worked carefully and doggedly on these sources,
and FR IIs are described primarily in
\citet{2000A&A...358..104C,2002ApJ...571..247C,2002A&A...394..791C}. The entire
data set and analysis is consistent with (but preceded!) the inferences from
the infrared. \citeauthor{2002ApJ...571..247C} argue that the radio galaxies that fall on the
well-populated (putative) synchrotron line in the optical/core-radio plane
are likely to be nonthermal AGN. The group of radio galaxies with larger optical
flux than expected for synchrotron radiation then host a visible
Big Blue Bump/Broad Line Region.

These authors further suppose that any opaque tori would be larger than the
optical point sources, in which case they would block the optical light;
therefore there are no tori in most such cases. Although M87 is an FR I (or
hybrid) source, it's worth mentioning in this section that its HST point source
is indeed small, because it varies on timescales of months.\footnote{\setlength\baselineskip{1ex}
However, this in itself is still no proof that it's smaller than any possible torus.}

Detailed study of 100 low luminosity 3CR radio galaxies with HST at
$1.6\mu$ is quite consistent with the prior conclusions of the \citeauthor{2002ApJ...571..247C}
group. In particular the low ionization galaxies of both FR types show ``central
compact cores'' which are probably nonthermal in nature \citep{2010ApJ...725.2426B}.

Powerful supporting evidence by the same group comes in the form of a
spectroscopic survey of $z < 0.3$ galaxies \citep{2010A&A...509A...6B}.  As had
been noted by e.g., \citet{1979MNRAS.188..111H}, FR II radio galaxies can be
naturally divided into low and high ionization objects. But \citet{2010A&A...509A...6B}
go a big step further and assert that their emission-line ``excitation
index'' is bimodal (their Fig.~4)! The first and only other claim of a related
bimodality of which I am aware is that of \citet{2004MNRAS.351..733M}, who cite such
a feature in the distribution of $L/L_{\rm Edd}$ at a value of $\sim0.01$. It's
worth keeping that value in mind.

Just as for the infrared data, the demographic conclusion from these optical
studies is that the fraction of 3CR FR II radio galaxies with visible or
hidden quasars is relatively small at low redshift, increasing up to at least
$z \sim 0.6$ \citetext{see also \citealp{2004A&A...428..401V}, and for a
slightly different opinion, \citet{2010ApJ...722.1333D}}.

Finally, it's been found that as a group, ``radio loud'' galaxies and quasars
(in this case with a flux cutoff or 3.5mJy at 1.5GHz --- which is extremely low
compared with the 3CRs) are clustered differently, with the quasars favoring
richer environments \citep{2010MNRAS.407.1078D}. This result is consistent with the views
expressed here because objects with $L_\nu$ (1.5GHz) $\gtwid10^{33}$
erg/sec/Hz (essentially the range of 3CR radio galaxies and
quasars) do cluster more like quasars, according to that paper. Nevertheless it
strongly suggests that the shadowing unification doesn't apply at lower radio
luminosities.\footnote{\setlength\baselineskip{1ex}
Another easy but worthwhile ``armchair ApJ Letter'' could be
written to test whether restriction of the \protect\citet{2010MNRAS.407.1078D}
radio galaxies to those of high radio luminosity would cause their clustering
properties to match those of the quasars.}

\subsection{X-rays\label{sec:classic-xray}}
Hidden radio quasars are characterized by large
($\sim10^{22}$ -- $10^{25}$ cm$^{-2}$ or more) cold absorbing columns, like the
Seyfert 2s and radio-quiet quasar 2s. The Low Ionization Galaxies generally don't show large
columns\footnote{\setlength\baselineskip{1ex}
One exception is found in \protect\citet{2011MNRAS.413.2358R}.
For a radio-quiet exception, see \citet{1984ApJ...285..458F}.} and this suggests
that they lack tori and Broad Line Regions, although it's always possible
that the jet continuum emission extends beyond a torus \citep{2009MNRAS.396.1929H}.
See their Fig.~16 and their table 7 for the
columns for Low Ionization Galaxies and other classes of AGN. These authors
argue persuasively that one can separate X-ray components from hidden quasars
and from jet emission, modeling the X-ray spectra with high-column and zero-column
components, respectively (see Figure~10).

Thus with reasonable SNR in the X-ray spectrum, one can say which spectral
component likely dominates that ostensibly derived from jet synchrotron emission
and those directly related to a copious accretion flow --- and often one
dominates completely. Examples of each type are shown in the FR II
galaxies in \citet{2005ApJ...621..167R}; \citet{2007ApJ...659.1008K};
\citet{2007ASPC..373..223T}; and
\citet{2010ApJ...710..859E}. The X-ray spectroscopic survey
of high-$z$ 3CR objects of \citet{2009cfdd.confE.206W}, like the infrared study
of \citet{2010ApJ...717..766L}, is extremely supportive of complete unification
of 3CR radio galaxies and quasars at $z\gtwid1$.

\begin{minipage}{5in}
\vskip 2.75truein
\includegraphics{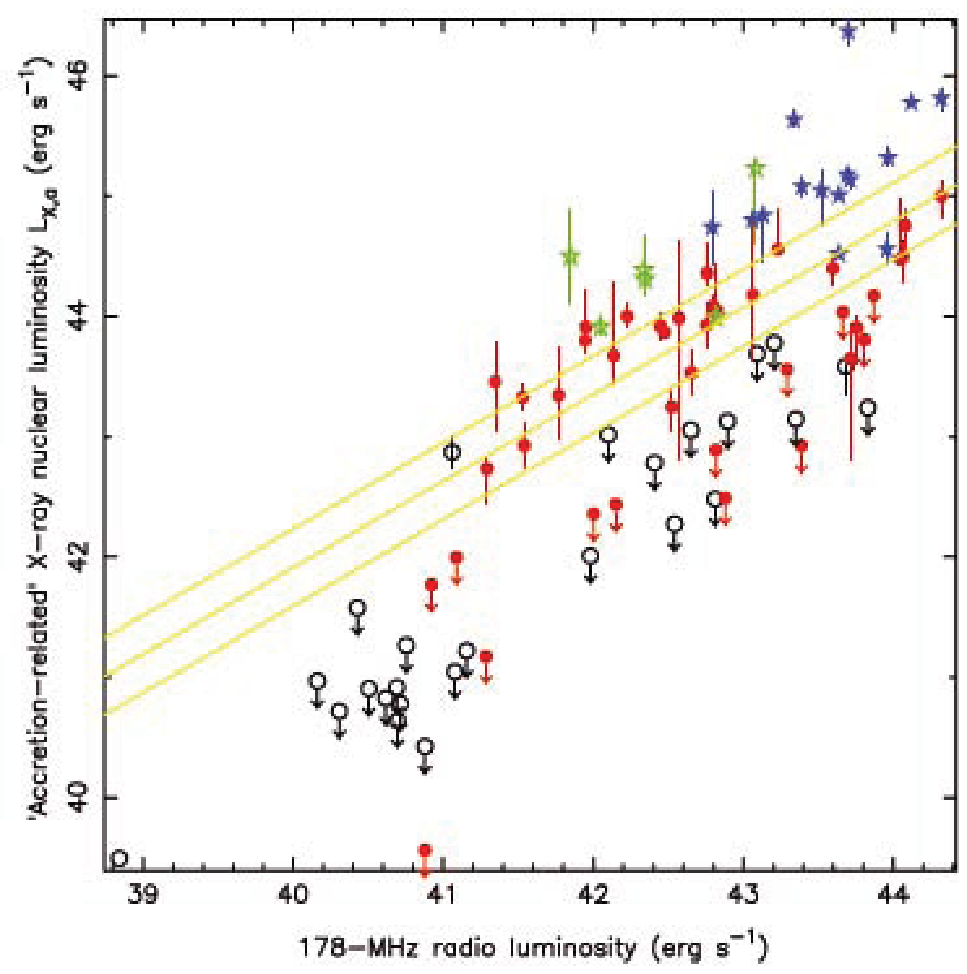}
\setlength{\baselineskip}{1ex}
\vskip 4mm
{Figure 10. The vertical axis shows the X-ray luminosity for the `accretion-related'
(absorbed) component as a function of 178-MHz total radio luminosity for the
$z < 1.0$ 3CRR sample. Regression is for detected Narrow Line Radio Galaxies
only. Black open circles indicate Low-Ionization Narrow Line Radio Galaxies,
red filled circles High-Ionization Narrow Line Radio Galaxies, green open stars
Broad Line Radio Galaxies and blue filled stars quasars. Clearly the Low
Ionization objects have little or no thermal emission. \citetext{Adapted from
\protect\citet{2009MNRAS.396.1929H}.}}
\end{minipage}
\vskip 3mm

The putative accretion-disk K-$\alpha$ lines in the thermal radio galaxies and
quasars seem to be weaker and narrower than in Seyferts, a fact often attributed
to an inner edge of an optically thick flat accretion disk at a greater radius
than that at which the disks in radio quiets terminate \citetext{e.g. \citealp{2005ApJ...618..139O,
2009ApJ...700.1473S,2010ApJ...721.1340T}}. The accretion disks are often taken
to have inner edges at the Innermost Stable Circular Orbit\footnote{\setlength\baselineskip{1ex}
An interesting possibility is that a large ISCO results from disk counter-rotation
\protect\citep{2010MNRAS.406..975G}.}, but this is controversial \citetext{e.g.
\citet{2000ApJ...528..161A}}. If the innermost part of the opaque thin disk is
missing, the efficiency of the standard\footnote{\setlength\baselineskip{1ex}
Optically thick, geometrical
thin disks with the alpha viscosity prescription, e.g., \protect\citet{1981ARA&A..19..137P}.}
disk is reduced, and also the thermal spectrum is cooler, so one would expect
(but doesn't see) corresponding changes in the spectral line ratios and
luminosities \citep{2005ApJ...618..139O}. Also in a recent study by \citet{2008MNRAS.390.1217M}
on a sample selected at 20--40 keV for their six ``definite'' Broad Line Radio
Galaxies, ``we only find marginal evidence for weaker reprocessing
features in our objects compared to their radio quiet counterparts.''
More commentary on this can be found in \citet{2010PASJ...62.1177K}.

\section{FR I RADIO GALAXIES}
\subsection{Radio Properties}
These lower-luminosity ($\ltwid 2 \times 10^{32}$ erg/s Hz$^{-1}$ at 1.5GHz)
big radio doubles generally have fairly symmetric twin jets on $>1$kpc scale,
and the lobes are edge-darkened with no terminal hotspots.
Much of the VLBI data on FR Is (and much of it on low-luminosity FR IIs) come
from the group behind these references: \citet{2001ApJ...552..508G,2005ApJ...618..635G};
T.~Venturi, pc, 2010. The data, though somewhat sparse on speeds especially,
are consistent with the assertion that FR I jets start out relativistic, 
with the FR Is being a little slower than the low-luminosity FR IIs. The
fraction of sources studied in the isotropically selected 2005 sample which are
visibly two-sided on VLBI maps is $\sim30$\%, vs.\ 5--10\% in earlier core-flux-selected
samples. Somewhat of an update was provided by \citet{2009A&A...505..509L}, with
single-epoch data on low-frequency selected FR Is, with aggregate results
consistent with a single unified (beam) model.

The overall statistics on the depolarization asymmetry, another powerful
constraint on the inclination distribution, show that the effect is weaker than
for quasars, and probably consistent with an isotropic distribution, though data
are scarce. For example, \citet{1997A&A...326..919M} looked at this for an FR I sample, and found
that the depolarization asymmetry is usually weak. \citet{1996IAUS..175..397G}
had come to a similar conclusion, noting however that strongly one-sided
jet sources have strong depolarization asymmetry. \citet{1995A&A...300..643C} found a
strong depolarization asymmetry in 2 out of 3 intermediate radio luminosity
(between FR I and FR II) radio galaxies. I think that more depolarization work
should be done, not just for the AGN field but for understanding galaxy and
cluster hot gas atmospheres, where the depolarization presumably takes place.

\subsection{Infrared}
Finally I'm getting to a topic that is a little bit controversial. It's not
{\it very} controversial in that everyone seems to agree that most FR Is have
predominantly nonthermal radio/infrared/optical and X-ray nuclei, e.g. \citet{2004A&A...426L..29M}.
But it is important not to overgeneralize, and assert a direct
connection between FR I morphology and a nonthermal engine.\footnote{\setlength\baselineskip{1ex}
It is easy to imagine
that the few exceptions can be attributed to the grossly different timescales on
which the core ($\sim1$ yr) and the lobes ($\sim10^8$ yr) are created. On the
other hand, the host-dependence of the luminosity cutoff, strongly suggests
environmental factors also play a role \protect\citep{1994ASPC...54..319O}.} I listed a few
exceptions in \citet{2002apsp.conf..151A,2002ASPC..284..147A}, going back to the first FR I
{\it quasar}, reported in 1984 by \citeauthor{1984AJ.....89.1658G}. (I don't know why
the more common FR I Broad Line Radio Galaxies don't impress people equally, but
some extra cachet seems to attach to those of quasar optical/UV luminosity.)

Recently there have been several papers modeling individual FR I nuclei as
purely nonthermal, which I think have been contradicted by subsequent Spitzer
spectra. For example, consider NGC6251\footnote{\setlength\baselineskip{1ex}
This is the first object with
good evidence of a narrow line flux variation: \protect\citet{1984ApJ...278..499A}, Fig.~1.},
analyzed by \citet{2003ApJ...597..166C}, for which the spectral index over most of
the infrared is inferred to be $\sim-0.6$. \citet{2009ApJ...701..891L} shows that the
mid-IR spectrum exceeds their predictions, and shows strong dust emission
features. Our conclusion from the SED is that there is hot dust with at least
as much flux as that due to synchrotron radiation. The aperture was $\sim4^{\prime\prime}$
but that may not matter much because dust emitting at a few microns must
be near a somewhat powerful optical/UV source, i.e., the nucleus. It'd be easy
to check from the ground. Thus the substantial near- and mid-IR dust emission
may signal a hidden thermal optical/UV also. See however
\citet{2008ApJ...678...78G}, who shows that the X-rays are likely to be
dominated by the jet in NGC6251. I think a problem with most published
decomposition of radio galaxy infrared spectra is that they require slopes
flatter than those of Blazars, opposite to the beaming prediction.

A Spitzer IRS spectrum for BL Lac is shown in Fig.~2 of \citet{2009ApJ...701..891L}:
the slope is $\sim-0.8$ between 5 and $30\mu$, considerably steeper than the synchrotron
slopes in most published nonthermal models for radio galaxies. What's more,
\citet{1988AJ.....95..307I} find that ``The spectra of Blazars steepen continuously
between $10^9$--$10^{15}$ Hz\dots the [frequency] at which the energy
distribution turns down in $\sim 2 \times 10^{11}$ Hz with a very narrow range
of spectral indices. Half of the Blazars with less than $10^{11}$ L$_\odot$ show
evidence for thermal infrared components\dots'' to which I'd add: which should be
much more conspicuous and thus more widespread in the high-inclination objects
(radio galaxies). Moreover, ``The average Blazar spectrum is flat ($\alpha\sim 0$) at $10^9$ Hz
and steepens continuously to $\alpha\sim -1.5$ at $10^{15}$ Hz. Table 4 shows
that the infrared slopes from $3\times 10^{12}$ Hz ($100\mu$) to $3\times10^{14}$
Hz ($1\mu$) are {\it all} steeper than 1.'' For the SEDs of FR I synchrotron
components, with lots of infrared Spitzer data, and both with and without dust
bumps, see \citet{2009ApJ...701..891L}: in our models, which feature relatively low
synchrotron contributions throughout the infrared, the slopes are steeper than
those found by other investigators and thus more reasonable in my opinion.

There are at least two ways around this objection to the required flatness
of the synchrotron components in some of the
published infrared decompositions. First, the emission from high-inclination may
be dominated by a slow-moving component which is for some reason intrinsically
flatter, and not directly related to the strong beamed component \citetext{e.g.
\citealp{2000A&A...358..104C}}. Also, the Blazar samples aren't necessarily matched to the
lobe-dominated radio galaxy samples and this could conceivably make a difference.
They do, however, include many objects with FR I diffuse radio power.

\Citet{2004IAUS..222...75V} account for most of the nonstellar radiation from 3CR270
($=$NGC4261) with a nonthermal model, but find some evidence for a weak thermal
component. Our Spitzer data show a big dust bump, which covers $3\mu$--$100\mu$,
and dominates the infrared energetically, at least as observed in the $4^{\prime\prime}$
Spitzer aperture. Please see Fig.~9 of \citet{2009ApJ...701..891L} for our spectral
decomposition, and the location of the synchrotron component. The Big Blue Bump is
extremely well correlated with the Broad Line Region in AGN, and I consider the possible
detection of broad polarized H-$\alpha$ in this object by \citet{1999ApJ...525..673B}
well worth following up. Again skipping ahead to the X-ray, \citet{2005ApJ...627..711Z}
conclude that 3CR270 is a heavily absorbed nucleus,
$N_H\sim8\times 10^{22}$~cm$^{-2}$, far higher than most FR Is \citetext{see Figure~10
here; also \citealp{2006A&A...451...35B}}. Synchrotron is thought by \citet{2005ApJ...627..711Z}
to contribute only $\sim10$\% of the X-ray flux.

The detailed discussion of Cen~A in \citet{2004ApJ...602..116W} still represents
our views on this controversial and somewhat complicated case. We think it
contains a hidden Big Blue Bump/Broad Line Region. Optical polarization imaging
is relevant for this FR I radio
galaxy. \citet{2007A&A...471..137C} have measured the percent polarization of the HST
nuclear sources at $\sim6060$\AA\ in several FR I galaxies. Restricting to those
with PA errors $\le10^\circ$ (the error functions have strong tails, unlike the
Gaussian function), the seven remaining objects are a few percent polarized at
random-looking angles. Cen~A has a similarly puzzling optical (R/I band)
polarization, influenced greatly by a foreground dust lane \citep{1996ApJ...459..535S}.

In the near-IR K band however, one can sometimes see through kpc-scale dust
lanes of modest optical depth to the nuclear occultation/reflection region
\citetext{\citealp{1990ApJ...363L..17A,2004ApJ...602..116W}; see also
\citealp{1986Natur.322..150B}}. \citet{1996MNRAS.278..406P} report on both the near-IR polarization
and the millimeter polarization (which turns out to be crucial): the
polarization of the nucleus after various corrections is given as an impressive
17\% (``in the near-IR''), and exactly {\it perpendicular} to the inner
radio axis. This is expected for hidden thermal AGN rather than for Blazars.
\citetext{Refined values can be found in \citealp{2000ApJ...544..269C}.} \citeauthor{1996MNRAS.278..406P} 
remark that the millimeter polarization, given simply as ``zero,'' is ``not\dots
consistent\dots with that of BL Lacs.'' As noted, we believe that near-IR
observations often see through the
dust lanes, enabling us to see this very high polarization exactly
perpendicular to the radio jet, as we demonstrated with the radio galaxy
2C223.1 \citetext{\citealp{1990ApJ...363L..17A}; we used a 1-channel polarimeter [!]
but our measurement was accurately confirmed with a modern instrument}.

The mere fact that the PA is constant in time for each near-IR observation of
Cen~A is unlike BL Lacs (or compact synchrotron sources in general), as
is the perpendicular relation to the radio jet. We also think that the spatially
resolved azimuthal off nuclear near-IR polarization \citep{2000ApJ...544..269C} is
most consistent with scattering from a normal Type 1 nucleus.

For Cen~A, we mention the X-ray spectrum here \citep{2007ApJ...665..209M}.
The superb Suzaku spectrum shows a column above $10^{23}$ cm$^{-2}$ for two
separate components, and many narrow fluorescent lines, including
Fe K-$\alpha$, like a Seyfert 2.

Going back to the mid-IR data on FR Is and FR IIs generally, an imaging survey of nearby
objects at $12\mu$ by \citet{2010A&A...511A..64V} revealed results which are
generally understandable and consistent with other arguments: the broad line
objects, all FR IIs, were easily detected at 7 mJy sensitivity (they quote $10
\sigma$!), as well as most of the High-Ionization Narrow Line Objects (also FR
IIs). The low-excitation galaxies of both types were not detected.

Spitzer is much more sensitive than any ground-based instrument. The current
state of the mid-IR art survey of FR Is from the IRS spectrograph is described
in \citet{2009ApJ...701..891L}. Here's where there is a little more controversy.
We observed 25 FR I radio galaxies, and carefully removed the star formation
contributions as well as possible using the PAH features, and also removed old
stellar populations using the Rayleigh-Jeans tail of the starlight, and using
the AGB star features at longer wavelengths. We reached the following conclusions
for the 15 putative pure-synchrotron sources described in \citet{1999A&A...349...77C}.
Of the 15 sources with ``optical compact cores'' from the Chiaberge group and
others (see the ``Optical'' section below), we see four with the infrared dominated
by contributions from the host galaxies. In another four of the galaxies with
optical point sources (but probably no exposed Big Blue Bump/Broad Line Region),
warm dust emission dominates, and is probably at least in part due to hidden nuclei,
contrary to the conclusions from the optical papers. In seven cases, synchrotron
radiation dominates the mid-IR. The comparison to the \citeauthor{1999A&A...349...77C} core
decompositions cannot be considered definitive however because of the larger
Spitzer aperture.

\subsection{Optical}
Some information about the optical point sources was used above to provide
context for the IR fluxes, but we must note here that these cores
\citep{1995ApJ...448..521Z,2003AJ....125.1795Z,2002AJ....123.1334V,1999A&A...349...77C,2000A&A...358..104C}
have been used to argue for synchrotron optical emission and no powerful hidden AGN
or tori in most FR I radio galaxies; \citet{2010ApJ...725.2426B} is closely related.
\citet{1998ApJS..114..177Z,2003AJ....125.1795Z} find that the
low-luminosity radio galaxies with detected central compact optical cores
are the ones with visible (single or highly one-sided) jets, and thus probably
low inclinations. This is compatible with the jet idea for the optical cases.

There has also been a series of papers by the Chiaberge, Capetti group
\citep{2005A&A...439..935C} on emission lines from low-luminosity radio galaxies.
These papers report that the putative synchrotron continuum is sufficient
to produce the observed emission lines, if the covering factors average
about 0.3.  They estimate the ionizing continuum from that in the UV with
power laws.

Another ambiguity is that the UV continuum can be very steep, and the
reddening corrections may be very large and uncertain \citetext{\citealp{2002ApJ...571..247C}}.
Also the observed ``UV excess" in at least a few low luminosity
radio galaxies is due to starlight, according to \citet{2004MNRAS.347..771W}. In that
case too the ionizing radiation can't be quantified easily.

\citet{2005A&A...439..935C} also made a factor of 5 correction downward to estimate
the H-$\alpha$ emission line flux from the measured flux of the blend with the
[N II] doublet. The factor of 5 seems too high to me, and comes not from typical
AGN behavior, but from a UGC sample of ordinary LINERs. Also, there was
apparently no starlight subtraction before this factor was determined from
spectroscopy \citep{2003ApJS..148..419N}. \citetext{The effect of this can be estimated
from formulae in \citealp{1983ApJ...269..466K}.} Finally, they detected in
{\it most} cases probable broad bases to the H-$\alpha$ line, but not the
forbidden lines. These lines are said to be ``compatible with the broad lines
seen in LINERs'' by \citet{1997ApJS..112..391H}. Any Big Blue Bump accompanying
these lines would probably be undetectable, at least with present data, so this
is consistent with (but not proof of) the presence of a Big Blue Bump, albeit
of low luminosity. Ho (\citeyear{1999ApJ...516..672H,2009ApJ...699..626H}) has
argued however that for his LINERs with weak broad H-$\alpha$, the central
engines are radiatively inefficient. My overall conclusion regarding the central
compact optical cores is that they are indeed mostly synchrotron sources, but I
don't share the same degree of confidence as the various authors.

Let us now return briefly to the question of broad emission lines in FR I galaxies,
concentrating on AGN that can be observed with good contrast relative to the host
galaxies. The most familiar object of this type is 3C120, with a fast superluminal
VLBI source. Several others \citetext{\citealp{2002apsp.conf..151A} gives a brief compilation} are also
highly core dominated, including BL Lac itself, in which the broad lines need to
compete with the beamed radiation in order to be detected. This suggests that a
Broad Line Region is sometimes visible in an FR I radio galaxy, when seen at low inclination.
\citet{1995A&A...298..395F} wrote a clever paper on this ``missing FR I quasar population.''

\subsection{X-rays}
This section will be brief because the results of many excellent studies are
simple and consistent, within the noise and the limited number of sources
analyzed, and the selection biases specific to each. Refer again to the present
Fig.~10 \citep{2009MNRAS.396.1929H} for strong evidence of very weak (ostensibly)
accretion-related power Low-Ionization Galaxies, including both FR types.
Other papers are generally very supportive of (and in some ways anticipated)
\citet{2009MNRAS.396.1929H}.

Some recent surveys with lots of FR I results: \citet{2004ApJ...617..915D}; \citet{2006A&A...451...35B};
\citet{2005ApJ...621..167R}; \citet{2006ApJ...642...96E}; and \citet{2006MNRAS.370.1893H}.
Overall, the great majority of X-ray spectra of FR I radio galaxies suggest
nonthermal emission. This finds strong independent support in that most of these
objects (the current Fig.~10) differ from Cen~A, and do {\it not} show
the high absorption columns typical of hidden AGN.

\section{SMALL SOURCES: COMPACT STEEP SPECTRUM AND GIGAHERTZ-PEAK SPECTRUM}
The radio properties of these young sources are described briefly in Section 1,
the Introduction. The classic complete review is \citet{1998PASP..110..493O},
while a shorter but recent review is \citet{2009AN....330..120F}.

Recall that few can grow into large bright sources, because they are
nearly as common as the big ones (as selected in the centimeter region),
but have very short kinematic and synchrotron-aging lifetimes. Recall
also that many of the tiny kinematic ages probably measure just the
age of the current stage of activity, which may repeat many times. It's also
possible that their birthrate is extremely high, but that most fade out before
they grow.

The experts generally seem to agree that the radio-galaxy/quasar unification
holds fairly generally, for the well-studied very radio luminous population.
That is, the radio galaxies are in the thermal class. At more modest
luminosities, there is less information and some hints from the infrared that
this may not be the case. If so, they behave similarly to the giant doubles.

\subsection{Radio Properties}
\citet{2001MNRAS.321...37S} provide several good arguments for small ages and
unification by geometry. On the former, kinematic and synchrotron ages
are small and generally consistent. On the latter, the authors note that the
quasars are more core-dominant in the radio, and they have more
asymmetric morphologies consistent with oppositely directed twin jets. Several
optical and radio papers present evidence for absorption by molecules and
HI, preferentially for the galaxies and thus near the plane according to
the Unified Model, e.g., \citet{2002ApJ...568..592B}; \citet{2006MNRAS.370..738G}; and
\citet{2009AN....330..120F}.

\subsection{Infrared}
Astronomers studied this class of radio source in the infrared with IRAS
\citep{1994ApJ...428...65H} and with ISO \citetext{e.g., \citealp{2000A&A...358..499F}}.
These radio emitters show
generally high (quasar-like) power in the aggregate. The ISO mission was able
to make many individual detections, but the \citet{2000A&A...358..499F} paper did
not make a comparison of their observed galaxies with GPS/CSS quasars.

In the Spitzer era, \citet{2010ApJ...713.1393W} presented data for eight
{\it relatively radio-faint} ``compact symmetric objects,''
which heavily overlap the GPS class. Only one was a broad-line object (OQ 208);
one was a BL Lac Object. Their Fig.~8 shows a plot of the Si strength
vs.\ equivalent width of the $6.2\mu$ PAH feature, demonstrating that
hidden AGN (marked by moderate Si absorption and fairly weak PAHs) probably
dominate the mid-IR emission of the galaxies in all cases. The quasar has Si
slightly in emission,
also as expected for the unified model. However, if considered bolometrically,
the AGN luminosities are low (except for the quasar) and PAH features indicate
that star formation may contribute significant luminosity. Ionization levels
are low. The authors favor Bondi accretion or black-hole spin energy for most
of the galaxies, not a thermal Big Blue Bump, so in our parlance they would fall
into the non-thermal class.\footnote{\setlength\baselineskip{1ex}
This refers to the overall SED, and not the infrared emission specifically.}

Our larger Spitzer survey \citep{2010ApJ...724.1193O} contains 13 quasars and
11 radio galaxies from the 3CR catalog. It contains objects of substantially
higher redshift (0.4--1.0) and luminosity ($10^{34}$--$10^{35}$ erg
s$^{-1}$ Hz$^{-1}$ at 1.5 GHz and 5 GHz), as compared with the
\citet{2010ApJ...713.1393W} sample.

The radio luminosities of our sample sound much higher than most FR IIs, where
the FR I/II cutoff is $\sim2 \times 10^{32}$ erg sec$^{-1}$ Hz$^{-1}$
at 1.5GHz, but the
CSS and especially the GPS sources are much weaker relative to the big
doubles at low frequency. Also note that this and their small sizes
indicate that they contain much less energy in particles and fields overall.

It is not so easy to get a complete isotropic sample of GPS sources
because they don't have dominant isotropic lobe emission. Possibly
one could select by an emission line, preferably in the infrared, after
radio classification. We took the GPS sources from \citet{1998A&AS..131..303S},
a complete GHz-selected sample but likely with beaming effects
favoring low inclination. This might not be too bad since we are mainly
comparing quasars and radio galaxies, with the latter still at larger
inclinations by hypothesis (shadowing Unified Model); however they may
not be at the same level on the luminosity function. For the CSS
sources, we could find a roughly isotropic sample in the 3CR, and we took
them from \citet{1995A&A...302..317F}.

We find that the GPS/CSS galaxies in our sample are {\it all} powerful thermal
dust emitters, with vLv($15\mu$) $\sim5$--$500 \times 10^{43}$ erg s$^{-1}$.
The Si features behave somewhat erratically vs.\ optical type, although
most follow the pattern of emission for quasars and absorption for
galaxies. This can be accommodated by clumpy torus models, but since it's
specific to the small objects, it might also be from the effects of colder foreground
off-nuclear dust. The interpretation of the galaxies as hidden quasars
is greatly strengthened by the [Ne V] and [Ne VI] lines detected in all of
the quasars and many of the galaxies. We detect no PAH or H$_2$ features.

In summary, the infrared evidence favors a (nearly?) ubiquitous shadowing unified
model for the most radio-luminous small sources. Remember, however, these are
the highest-luminosity members of the class, and it's likely that at lower
luminosities, some radio galaxies lack the hidden quasars, based on
\citet{2010ApJ...713.1393W}.

\subsection{Optical}
Bright GPS and CSS radio sources tend to have strong line emission. All three
3CR CSS sources studied by \citet{2005A&A...436..493L}, two radio galaxies
and one quasar, show [O III]$\lambda$5007 narrow H$\beta > 10$, so they have
high ionization. There is evidence for shocks also in these nice HST spectra.

There is a wealth of information in \citet{1999ApJ...526...27D} and \citet{2000AJ....120.2284A}
on HST imaging of CSS radio galaxies. These groups imaged in the
H$\alpha$ or [O III] $\lambda$5007 line in tens of objects. Only the broad
line objects have point continuum sources, in accord with the Unified
Model. Thus there are almost no known ``bare synchrotron''
sources\footnote{\setlength\baselineskip{1ex}
An exception is PKS 0116+082, from \protect\citet{1997ApJ...484..193C}. This
really anomalous object has high and variable polarization like a Blazar
and good limits on broad H$\alpha$/$\lambda$5007, yet very strong narrow emission
lines. There is actually a
possible broad H$\alpha$ line in polarized flux, which is however
unexpected in a Blazar. This object is not analogous
to the much lower luminosity HST optical point sources studied by Zirbel
and Baum, and Chiaberge et al., and discussed in the previous two
sections. One point in common though is at least a few percent optical
polarization \protect\citep{2007A&A...471..137C}.} as described above for FR I and FR II
galaxies.  This is not really a demonstrated difference however since
until very recently, only the most luminous GPS/CSS sources have been
studied in detail. In fact, very recently \citet{2010MNRAS.408.2279K} have reported on line
emission on {\it fainter} small radio sources, finding that many have Low
Ionization Galaxies spectra, just like for the giant doubles.

Both \citet{1999ApJ...526...27D} and \citet{2000AJ....120.2284A} also found
that the line emission lies preferentially parallel to the radio axes, and
Axon et al. add that photon counting arguments require a hidden radiation source.
Those alone are powerful arguments for unification with quasars.

\subsection{X-rays}
There are several papers on X-rays from small radio sources.
\citet{2006A&A...446...87G} reported on a small sample of GPS galaxies at
redshifts between 0.2 and 1. All four with adequate SNR have large column
densities, ``consistent with that measured in High-Excitation FR II
galaxies,'' strongly indicating hidden quasars. The radio luminosities
are around $10^{34-35}$ erg s$^{-1}$ Hz$^{-1}$ at 1.5GHz, near the spectral
peaks. The 2--10 keV X-ray de-absorbed luminosities are $10^{44}$--$10^{45}$ erg
s$^{-1}$. (The authors give their $H_0 = 70$, but no other cosmological
parameters, so this is approximate.) Since that band covers only 0.7 dex
in frequency, the X-ray luminosity alone is $\gtwid 1 \times 10^{45}$ erg s$^{-1}$, and
the bolometric luminosity is likely to be at least a few times higher. There is
one exception, which is convincingly argued to be Compton-thick and so
opaque in the X-ray. Consistent results were reported by \citet{2008ApJ...684..811S}.

This group expanded their survey of GPS galaxies \citetext{\citet{2009A&A...501...89T};
neither study included quasars} and confirms and extends these results, again
concentrating on the most radio-luminous objects.

\section{Summary}
First a hearty congratulations to all the theorists who predicted that
accretion would become inefficient below $L\sim 0.01 L_{\rm Edd}$. Apparently,
the initial argument to this effect is that of \citet{1982Natur.295...17R}.
Parts of the theory were anticipated by \citet{1976ApJ...204..187S}
and \citet{1977ApJ...214..840I}. More recently, \citet{1997ApJ...489..865E}
quotes 0.4 $\alpha^2L_{\rm Edd}$ as the cutoff luminosity in
the $\alpha$ prescription, which they find fits well with black hole binary
state changes.

The first radio galaxy discovered and interferometrically mapped
was Cygnus~A, which has a whopping apparent brightness of 8700 Jansky
at 178~MHz. At low frequencies, you can actually point the VLA 90 degrees
away from Cygnus~A and map it in a sidelobe! (F.~Owen, pc 2010.)
Also interesting: considering the volume enclosed by the Cyg~A distance
and the very strong cosmological evolution of FR II radio sources, folklore
says only one universe in 10,000 should be so lucky as to have such a prize!

The first map of Cygnus~A is very impressive for the time, and the map
is very charming \citetext{Fig.~11 here, from \citealp{1953Natur.172..996J}}.
There is no question that it has a hidden quasar of moderate luminosity
\citetext{\citet{1994Natur.371..313A}; \citet{1997ApJ...482L..37O}; infrared fluxes from NED}.
See \citet{1996MNRAS.283L..45B} to find out how such a modest quasar can be so
incredibly powerful in the radio.

\begin{minipage}{5in}
\vskip 2.75truein
\includegraphics{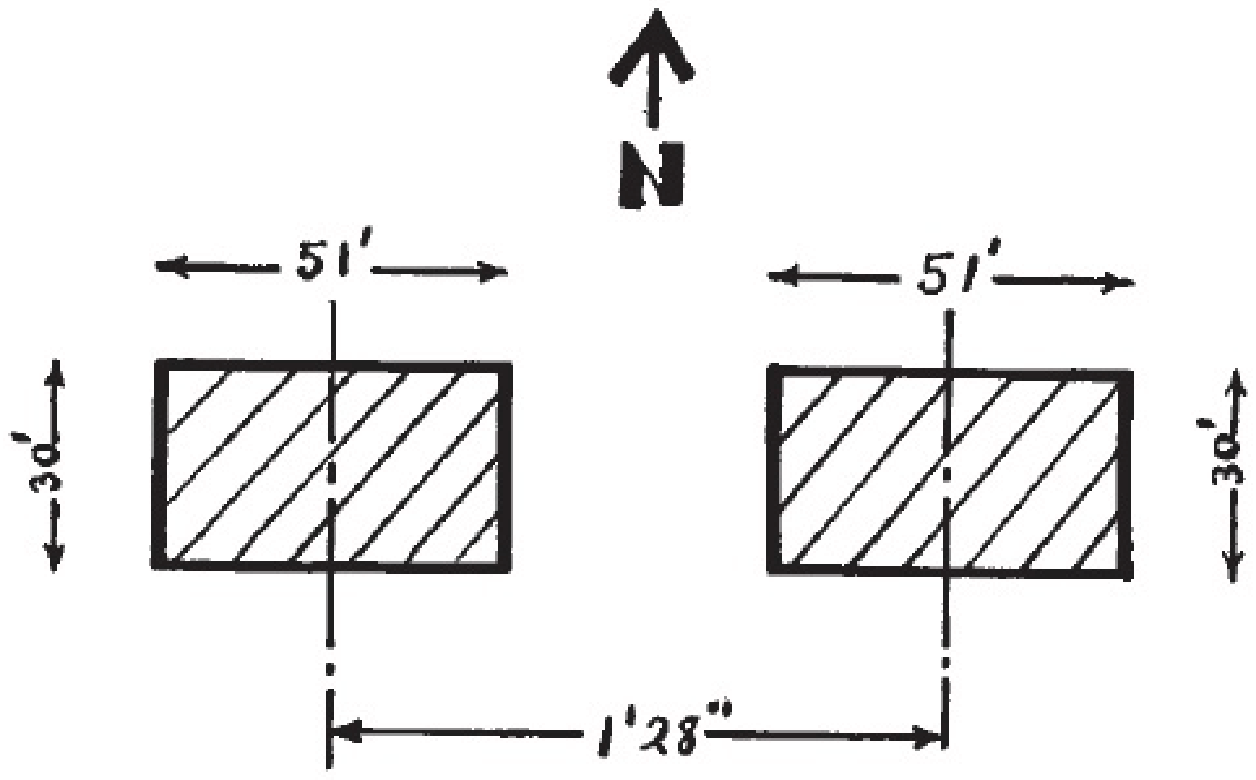}
\setlength{\baselineskip}{1ex}
{Figure 11. Approximate intensity distribution of the extra-terrestrial
radio source in Cygnus. \citep{1953Natur.172..996J}. This very early
interferometric observation first revealed the double-lobed nature of most
radio galaxies.}
\end{minipage}
\vskip 3mm

We know now that some powerful radio galaxies, and many weaker ones,
lack detectable visible or hidden thermal AGN. It is, however, very important
to remember that in {\it all} but a couple of contentious cases, Type 2 
radio quiet quasars have hidden AGN,
until you get to the LINER (very low luminosity) regime, which then shows ADAF
behavior according to most investigators \citetext{\citet{2006ApJ...648L.101E};
\citet{2000A&A...358..104C}; \citet{2009ApJ...699..626H}; many others},
along with weak broad wings to H$\alpha$ in many cases.

The time has come to stop proving this! We have altogether quite settled
the question. It's worthwhile exploring more parameter space, e.g., weaker
radio/IR/optical/X-ray sources, and to follow up on individual interesting
cases, but the overall pattern is clear now for all types of bright radio source.

So what should we do instead? The result of all this work is that we can now
hold many things constant while we vary just one thing, the tremendous
thermal emission
(Big Blue Bump). The two types must differ drastically in their structure
on relativistic scales. We are limited by our imaginations in how to take
advantage of this situation. My group is trying to determine how the
thermal/nonthermal states correlate with VLBI properties, i.e. jet
launching, collimation, and proper motion, to the extent that's possible
with current VLBI angular resolution, all while holding the large-scale structure
constant as far as we can discern. Another obvious observation, which
really requires next-generation X-ray telescopes to do well, is to
compare the reflection signatures of the two types. The putative accretion disk
K-$\alpha$ fluorescence line isn't expected to be so broad or strong if there
is no opaque accretion disk at small radii. Next, AGN feedback in galaxy evolution
might be mediated by radiation pressure on dust in many cases, or else
by PdV work or particles from jets and lobes, or other mechanisms. Here we can keep
everything the same (?) as far as the latter go, but turn off the radiation.
Does it make a difference? These new data might actually help with the
physics of AGN and of galaxies, and not just the astronomy of AGN.  Enjoy!

\section*{Acknowledgements}
Several astronomers have provided advice and unpublished data for this paper.
These include S.~Baum, M.~Begelman, P.~Best, O.~Blaes, S.~Buttliglione, M.~Chiaberge,
C.M.~Gaskell, M.~Gu, M.~Haas, L.~Ho, S.~Hoenig, W.~Keel, P.~Kharb, M.~Kishimoto,
M.~Kunert-Bajraszewska, C.~Leipski, R.~Maiolino, C.~O'Dea, F.~Owen, S.~Phinney,
D.J.~Saikia, A.~Singal, D.~Whysong and B.~Wills. C.~Leipski, S.~Hoenig, D.~Whysong and
P.~Ogle provided unpublished figures. P.~Ogle, S.~Willner, and R.~Barvainis
suffered the most (after the author), locating many errors and ambiguities.
To the extent that the text sounds literate, my editor/typist Debbie L.~Ceder deserves credit.

\bibliography{antonucci-refs}

\end{document}